\documentclass[sigconf]{acmart}
\usepackage{array}
\usepackage{comment}
\usepackage{subfigure}
\usepackage{multirow}
\usepackage{xcolor}
\usepackage{algorithm}
\usepackage{multicol}
\usepackage{arydshln}
\usepackage{booktabs}
\usepackage{balance}

\AtBeginDocument{%
  \providecommand\BibTeX{{%
    \normalfont B\kern-0.5em{\scshape i\kern-0.25em b}\kern-0.8em\TeX}}}

\setcopyright{acmcopyright}
\copyrightyear{2024}
\acmYear{2024}
\setcopyright{acmlicensed}\acmConference[WSDM '24]{Proceedings of the 17th ACM International Conference on Web Search and Data Mining}{March 4--8, 2024}{Merida, Mexico}
\acmBooktitle{Proceedings of the 17th ACM International Conference on Web Search and Data Mining (WSDM '24), March 4--8, 2024, Merida, Mexico}
\acmPrice{15.00}
\acmDOI{10.1145/3616855.3635810}
\acmISBN{979-8-4007-0371-3/24/03}

\begin{document}

\title{Calibration-compatible Listwise Distillation of Privileged Features for CTR Prediction}

\author{Xiaoqiang Gui}
\affiliation{
  \institution{School of Software, Shandong University}
  \city{Jinan}
  \country{China}
}
\email{x.q.gui@mail.sdu.edu.cn}

\author{Yueyao Cheng}
\author{Xiang-Rong Sheng}
\affiliation{
  \institution{Alibaba Group}
  \city{Beijing}
  \country{China}
}
\email{chengyueyao.cyy@taobao.com}
\email{xiangrong.sxr@taobao.com}

\author{Yunfeng Zhao}
\author{Guoxian Yu*}
\thanks{*Guoxian Yu and Shuguang Han are the corresponding authors.}
\affiliation{
  \institution{School of Software, Shandong University}
  \city{Jinan}
  \country{China}
}
\email{yunfengzhao@mail.sdu.edu.cn}
\email{gxyu@sdu.edu.cn}

\author{Shuguang Han*}
\affiliation{
  \institution{Alibaba Group}
  \city{Hangzhou}
  \country{China}
}
\email{shuguang.sh@taobao.com}

\author{Yuning Jiang}
\author{Jian Xu}
\affiliation{
  \institution{Alibaba Group}
  \city{Beijing}
  \country{China}
}
\email{mengzhu.jyn@taobao.com}
\email{xiyu.xj@taobao.com}

\author{Bo Zheng}
\affiliation{
  \institution{Alibaba Group}
  \city{Beijing}
  \country{China}
}
\email{bozheng@taobao.com}

\renewcommand{\shortauthors}{Xiaoqiang Gui et al.}

\begin{abstract}
In machine learning systems, privileged features refer to the features that are available during offline training but inaccessible for online serving. Previous studies have recognized the importance of privileged features and explored ways to tackle online-offline discrepancies.
A typical practice is privileged features distillation (PFD): train a teacher model using all features (including privileged ones) and then distill the knowledge from the teacher model using a student model (excluding the privileged features), which is then employed for online serving. In practice, the pointwise cross-entropy loss is often adopted for PFD. However, this loss is insufficient to distill the ranking ability for CTR prediction. First, it does not consider the non-i.i.d. characteristic of the data distribution, i.e., other items on the same page significantly impact the click probability of the candidate item. Second, it fails to consider the relative item order ranked by the teacher model's predictions, which is essential to distill the ranking ability. To address these issues, we first extend the pointwise-based PFD to the listwise-based PFD. We then define the calibration-compatible property of distillation loss and show that commonly used listwise losses do not satisfy this property when employed as distillation loss, thus compromising the model's calibration ability, which is another important measure for CTR prediction. To tackle this dilemma, we propose Calibration-compatible LIstwise Distillation (CLID), which employs carefully-designed listwise distillation loss to achieve better ranking ability than the pointwise-based PFD while preserving the model's calibration ability. We theoretically prove it is calibration-compatible. Extensive experiments on public datasets and a production dataset collected from the display advertising system of Alibaba further demonstrate the effectiveness of CLID.
\end{abstract}

\ccsdesc[500]{Information systems~Recommendation systems}

\keywords{CTR Prediction, Privileged Features, Calibration-compatible Listwise Distillation}



\maketitle

\section{Introduction}

Click-through rate (CTR) prediction is an essential component in online recommendation systems~\cite{zhang2022keep,gao2023rec4ad}. While receiving a user request, a recommendation system retrieves a set of candidate items, ranks them, and then displays them to users. In the ranking stage, a CTR prediction model typically takes the user's features and candidate items' features as input. The model then predicts the user's probability of clicking the candidate item. In most cases, the same set of features are adopted in offline training and online serving to guarantee model consistency. 

\begin{figure}[!tbp]
\centering
\includegraphics[width=7cm]{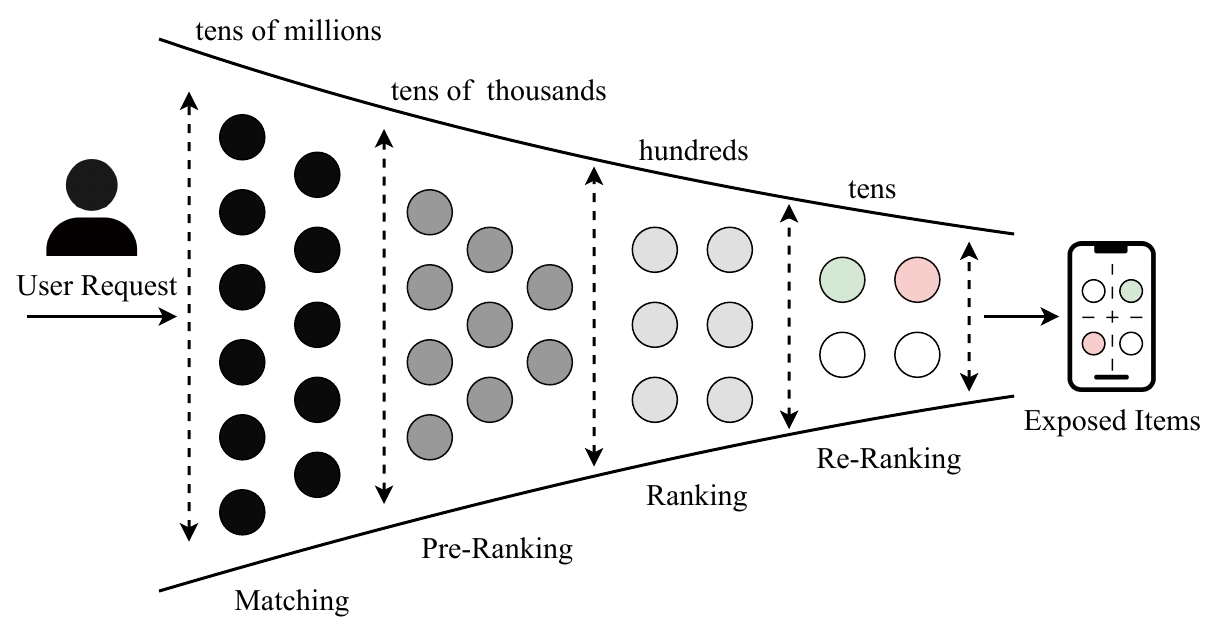}
\vspace{-1em}
\caption{An illustration of the formation process of the items a user sees
when visiting the online recommendation system.}
\vspace{-1em}
\label{rs_structure}
\end{figure}

However, previous studies have also identified that despite being useful during offline training, some features are not readily available for online serving~\cite{vapnik2015learning, xu2020privileged,yang2022toward}. For example, when predicting the probability of purchase after a user clicks the item\cite{gu2021real,chan2023capturing,zhao2023entire}, the dwell time on the detailed item page is a useful post-event feature that only exists in the offline training data because the online model cannot know this time before the user leaves the page. For ease of differentiation, we call the features that exist only during training \textit{privileged features} and those that exist during both training and serving \textit{non-privileged features}. There exist some informative privileged features in the CTR prediction task. Fig. \ref{rs_structure} illustrates the formation process of the items a user sees when visiting the online recommendation system. 
In the ranking stage, the CTR model retrieves hundreds of candidate items and generates tens of items that will be refined in the re-ranking stage and then exposed to the user. Therefore, the ranking model (i.e., the CTR model in the ranking stage) cannot observe exposed items before generating recommendations. That is to say, the features regarding the items displayed together with the candidate item on the same page are privileged ones for the ranking model when predicting the user's probability of clicking the candidate item. In the following, we refer to these items and their features as \textit{contextual items} and \textit{contextual features}. Contextual items significantly affect the user click propensity of the candidate item.
Therefore, properly utilizing privileged contextual features for the ranking model can significantly improve CTR prediction performance. Note that contextual features are non-privileged features for the model in the re-ranking stage where there is no issue of training-serving inconsistency, and many attempts~\cite{ai2018learning,ai2019learning,pasumarthi2020permutation, zheng2022cbr} have been made.

To utilize privileged features, previous works ~\cite{guo2019pal,zhao2019recommending,zhang2023towards} have examined adding them into the detached shallow tower of the prediction model and discarding them at serving. 
However, in this way, the input distributions at the training and inference (i.e., online serving) phases are inconsistent, i.e., the covariate shift issue still exists, which usually hurts the model generalizability~\cite{SugiyamaKM2007CovariateShiftAdapt}. To preserve the offline-online consistency while incorporating the privileged features, ~\citet{xu2020privileged} proposed a privileged features distillation (PFD) framework. PFD introduces student and teacher models, in which the teacher model utilizes both privileged and non-privileged features for better training performance. In contrast, the student model only trains on top of non-privileged features and acts as the final model for serving. The knowledge learned from the teacher model (i.e., the teacher model's predictions) is employed as soft labels to instruct the student model. 

Knowledge distillation of existing PFD methods \cite{xu2020privileged,liu2022position,yang2022toward} is typically implemented with a pointwise loss. Despite being effective, the pointwise distillation loss separately treats each item based on the identical and independently distributed (i.i.d.) assumption, which is against reality since items in a recommendation system are often displayed in a list format, and the probability of clicking on an item is affected by other items on the same page \cite{cao2007learning,xia2008listwise}. Consequently, the pointwise loss fails to consider the information of the relative item order ranked by the teacher model's predicted click probabilities (pCTR) for items within the same page, making it insufficient to distill the ranking ability from the teacher model.



To address the above issue, a straightforward approach is to extend the PFD framework with the listwise loss as the distillation loss. The listwise loss treats each list as an optimization instance and naturally accounts for the non-i.i.d. nature of items' pCTR. Indeed, we empirically observe an improved ranking performance for the listwise-based PFD methods over the pointwise-based one. However, when employed as distillation loss, the commonly used listwise losses can destroy the probabilistic meaning of the student model's predictions as pCTR, degrading the student model's calibration ability. For the CTR prediction task, the calibration ability, i.e., \textit{whether or not the predicted click probability aligns with the actual click-through rate}, is another important factor for measuring the model performance~\cite{chaudhuri2017ranking,liu2022position,yan2022scale,bai2023regression,sheng2022joint,zhao2023copr,wu2022adversarial}. 
For example, considering the cost-per-click (CPC) system in online advertising, a candidate advertisement (ad) is ranked and charged by the strategy of effective cost per mile impressions (eCPM), which is computed as $ \text{eCPM} = 1000 \cdot pCTR \cdot \text{Bid}_{\text{CPC}}$, where $\text{Bid}_{\text{CPC}}$ is the bid price from the advertiser, and $pCTR$ denotes the predicted click-through rate (predicted click probability) of the ad, which the predictive model produces. The computation of eCPM shows that mis-calibrated pCTR can damage the user experience, prevent advertisers from achieving their marketing goals, and eventually affect the revenue of advertising platforms. Thus, a well-calibrated prediction is the key to bidding, charging, and ranking the candidate ads. 

To improve the model's ranking ability while preserving its calibration ability with the listwise distillation loss, we propose a Calibration-compatible LIstwise Distillation (CLID) approach for CTR prediction. In particular, inspired by RCR \cite{bai2023regression}, a calibration-compatible listwise loss in the field of Learning-To-Rank (LTR), we first define the \textit{calibration-compatible} property of the distillation loss within the PFD context: \textit{the distillation loss (using the soft label from the teacher model) can achieve the global minima simultaneously with the losses (using the click label) of the student and teacher model}. We show that widely employed listwise losses (e.g., ListNet \cite{cao2007learning}, and ListMLE \cite{xia2008listwise}) are not calibration-compatible when applied as the distillation loss. 
Subsequently, to achieve the calibration-compatible listwise distillation loss, we carefully design the listwise loss by exploiting the cross-entropy to measure the discrepancy between two relative item orders on the same page, encoded using normalized pCTR generated by the student and teacher models, respectively. Finally, we provide theoretical proof that the tailored listwise loss in CLID is indeed \textit{calibration-compatible}. 

Extensive experiments over various baselines on two public datasets and a production dataset collected from the display advertising system of Alibaba verify that CLID achieves substantial ranking performance improvements while preserving the model's calibration performance. The main contributions of this research are as follows:\\
\noindent (i) We observe that the popular pointwise-based PFD methods disregard the non-i.i.d. nature of data distribution for CTR prediction and suggest extending the distillation loss with listwise ones to improve the model's ranking ability better. Moreover, we further show the widely employed listwise losses can compromise the model's calibration ability, a crucial aspect in CTR prediction, rendering them unsuitable for direct application in PFD.\\
\noindent (ii) We propose a Calibration-compatible LIstwise Distillation (CLID) approach that carefully designs the listwise distillation loss to distill the ranking ability from the teacher model without destroying the model's calibration ability. Besides, we theoretically prove the tailored listwise distillation loss satisfies the defined \textit{calibration-compatible} property, thus yielding well-calibrated predictions.\\
\noindent (iii) We validate the effectiveness of CLID on both public and production datasets. Experimental results show that CLID significantly improves the student model's ranking ability while preserving its calibration ability.

\section{Related Work}
\textbf{Privileged features}~\cite{lopez2016unifying} are available during training but inaccessible at the testing time due to the expensive real-time computation overhead or feature unavailability before online inference. 
Privileged features widely exist in various machine learning applications, including the emotion recognition~\cite{wang2015emotion}, action detection ~\cite{luo2018graph}, image super-resolution ~\cite{lee2020learning}, etc. Several Unbiased Learning To Rank (ULTR) \cite{ai2021unbiased} methods have been made to utilize the privileged exposed position features of items for CTR prediction.
One line of work ~\cite{ling2017model,haldar2020improving, guo2019pal,zhao2019recommending, zhang2023towards} incorporates the position information as the feature of the detached shallow tower when training and adopts the default position or discards the external tower at inference. As a result, the input distributions of these methods during training and inference are inconsistent, which causes unstable improvements in the model's ranking ability and vast degradation of the model's calibration ability, as shown in our later experiments.
Instead of regarding the position as a feature, another line of work attempts to directly model the position influence with result randomization~\cite{hofmann2013reusing, swaminathan2015batch, wang2016learning} or Inverse Propensity Weighting (IPW)~\cite{ai2018unbiased, chen2021adapting, wang2018position}. These approaches are often ad-hoc and cannot be simply extended to various privileged features, such as dwell time and contextual features. 
Note that the privileged contextual features here are different from the non-privileged "contextual features" mentioned in other works, which either focus on the "contextual features" in the re-ranking stage ~\cite{ai2018learning,ai2019learning,pei2019personalized,zheng2022cbr} or the "contextual features" of each item within the user's historical behaviors ~\cite{fan2022modeling,huang2022deep,li2023decision}. Therefore these works can not be compared with CLID when conducting experiments.
This paper employs the privileged knowledge distillation (PFD) framework to model various privileged features.

\textbf{Privileged features distillation} is a popular and powerful technique to exploit privileged features while solving the training-serving inconsistency. 
~\citet{xu2020privileged} proposes the Privileged Features Distillation (PFD) technique that feeds the teacher model with both non-privileged and privileged features and further demonstrates the superiority of PFD in the recommendation systems. 
~\citet{liu2022position} demonstrates that PFD achieves the state-of-the-art performance when utilizing the privileged position feature.
~\citet{yang2022toward} further analyzes the underlying mechanism of PFD in recommendation systems. However, these methods all employ the pointwise loss as the distillation loss that is insufficient to distill the teacher model's ranking ability for CTR prediction. 
Unlike them, our CLID designs the \textbf{calibration-compatible listwise distillation} loss to capture the inter-item dependency and distill the ranking ability while maintaining the model's calibration ability.

Our research is also closely related to the field of \textbf{Learning-To-Rank (LTR)}. In LTR, numerous attempts have been made to design effective loss functions, including pointwise, pairwise, and listwise losses~\cite{Wei2009Ranking,han2020learning}. Generally, the pointwise loss takes a single item as the learning instance~\cite{Li2007McRank,cossock2008statistical}, the pairwise loss views a pair of items as the learning instance~\cite{graepel2000large,burges2005learning}, and the listwise loss regards an individual list of items as a learning instance~\cite{cao2007learning,xia2008listwise}. 
Among these losses, the listwise loss naturally captures the non-i.i.d. data dependency and aims to learn a scoring function that minimizes the ranking loss and induces a perfect ranking list, thus producing better ranking performance than the pointwise and pairwise ones.
In addition, a listwise loss with a list size of $2$ degenerates into a pairwise loss.
Therefore, this paper mainly focuses on employing the listwise loss as the distillation loss.
However, the scores of commonly employed listwise losses (e.g., ListNet ~\cite{cao2007learning}, ListMLE ~\cite{xia2008listwise}, ApproxNDCG ~\cite{qin2010general}) are not well calibrated, limiting their usage in score-sensitive business applications, e.g., the online advertising platform. 
Recently, Bai et al. ~\cite{bai2023regression} proposed RCR which modified the mapping function in ListNet to align with various pointwise losses and proved that RCR and pointwise losses share the global minima in the context of LTR. Inspired by them, we carefully design the listwise distillation loss and prove that the loss satisfies the defined \textit{calibration-compatible} property within the PFD context. 


\begin{figure*}[h!tbp]
\centering
\includegraphics[width=.7\textwidth]{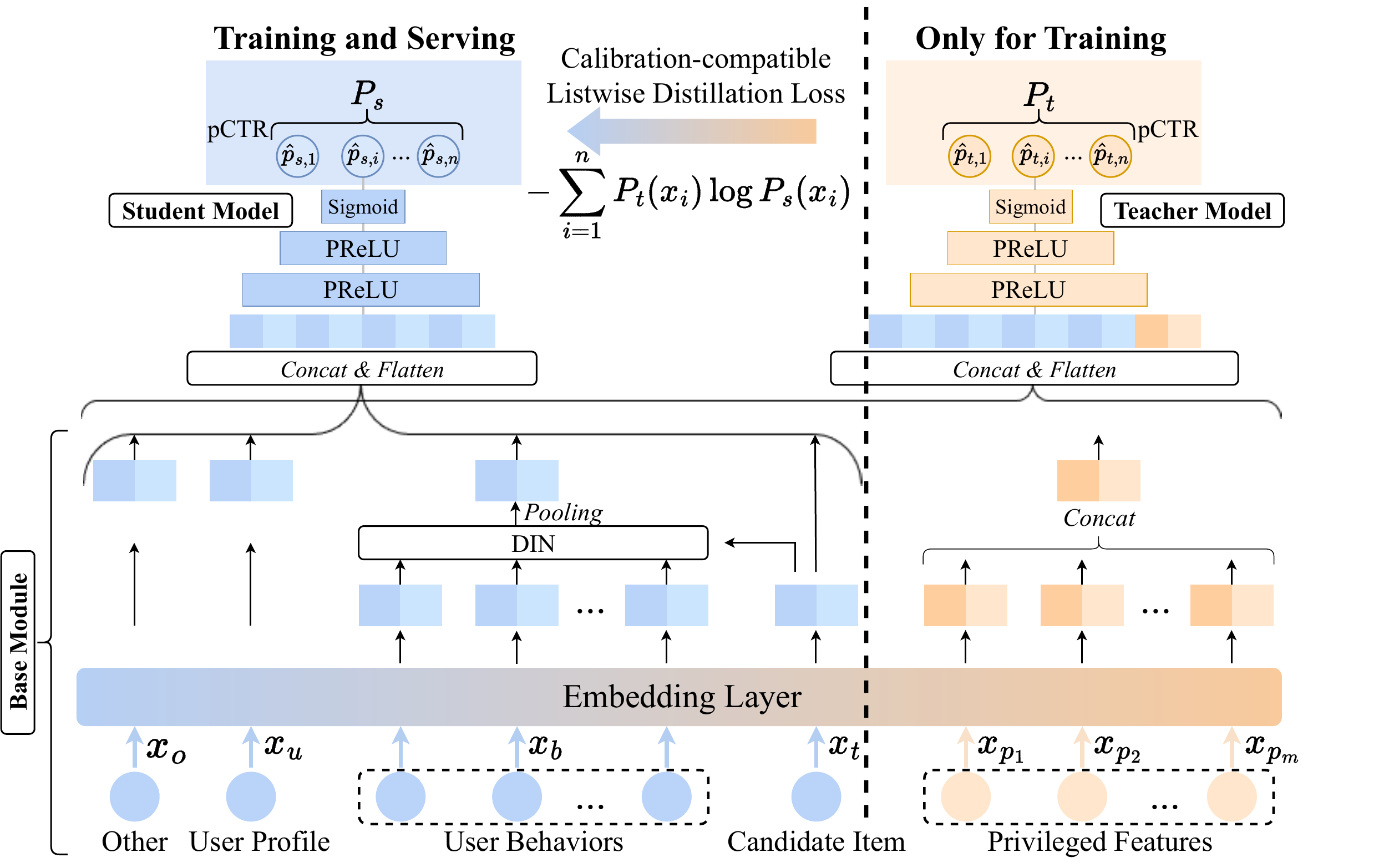}
\vspace{-1em}
\caption{The framework of our proposed CLID method. CLID consists of three parts: base module, student model, and teacher model. The base module takes non-privileged features (e.g., user profile and a set of features with regard to candidate item) and privileged features as input and maps them into the fixed-length vectors. Both non-privileged and privileged features are employed as the input of the teacher model for training, whereas only non-privileged features are employed as the input of the student model for training and serving. PReLU ~\cite{he2015delving} represents the commonly used activation function in CTR prediction.
The output pCTR of the student and teacher model is $\hat{p}_{s,i}$ and $\hat{p}_{t,i}$, respectively.
The calibration-compatible listwise distillation loss takes $\hat{p}_{t,i}$ as the soft label and instructs the learning of $\hat{p}_{s,i}$. Only the output pCTR of the student model is used for serving.}
\vspace{-1em}
\label{framework}
\end{figure*}

\section{Methodology}
In this section, we first give a brief overview of the system architecture and then introduce the proposed calibration-compatible listwise distillation (CLID) approach for CTR prediction.

\subsection{System Overview}
Figure \ref{framework} depicts the overall framework of CLID, which consists of three parts: a base module, a student model, and a teacher model. The base module first transforms discrete feature IDs into low-dimensional embeddings. Then those embeddings are aggregated to obtain a fixed-length vector. 
The teacher model utilizes the vector obtained from both non-privileged and privileged features as input, and outputs pCTR $\hat{p}_t$.
The student model shares the same MLP structure as the teacher model, while it only employs the vector obtained from non-privileged features as input and outputs pCTR $\hat{p}_s$.
In the meantime, the student model also adopts the calibration-compatible listwise distillation loss to distill the knowledge from $\hat{p}_t$. The final output $\hat{p}_s$ from the student model will be used for serving.

\subsection{Model Structure}
\label{3.2}
A typical CTR prediction task aims to predict the likelihood of a click for each user-item pair.
In the most common case, a CTR prediction model often utilizes non-privileged features, including
the user behavior sequence $x_{b}$, user profile $x_{u}$, candidate item $x_{t}$, and other features $x_{o}$, as input and outputs pCTR. 
\subsubsection{Base Module.} 
As shown in Figure \ref{framework}, the base module first utilizes the embedding layer to transform the high-dimensional features $x$ into low-dimensional embeddings $e(x)$. These features include non-privileged features and privileged ones. $m$ different privileged features are denoted as $x_{p_1},x_{p_2}, \dots, x_{p_m}$.
For user behavior features, the pooling layer is applied to transform the list of embedding vectors into a fixed-length vector. Before pooling, DIN ~\cite{zhou2018deep} is adopted for better modeling of feature interaction. With the concatenation layer, we can obtain the non-privileged representation vector $\mathbf{V}_r$ as the input of the student model: 
\begin{equation}
    \mathbf{V}_r = Concat\Big(e(x_{o}), e(x_{u}),e(x_{t}),Pool(\text{DIN}(e(x_b),e(x_t)))\Big)
    \label{V_r}
\end{equation}
We concatenate the low-dimensional embeddings of $m$ privileged features to obtain the privileged representation vector:
\begin{equation}
\mathbf{V}_p = Concat\Big(e(x_{p_1}),e(x_{p_2}),\dots,e(x_{p_m})\Big)
    \label{V_p}
\end{equation}
Since the privileged features are available at training but inaccessible during serving, we add them to the teacher model to guide the learning of the student model. The input vector $\mathbf{V}_t$ of the teacher model is composed of the non-privileged vector $\mathbf{V}_r$ and the privileged vector $\mathbf{V}_p$ as follows:
\begin{equation}
    \mathbf{V}_t = Concat(\mathbf{V}_r,\mathbf{V}_p)
    \label{V_t}
\end{equation}

\subsubsection{Teacher and Student Model.} 
Given a sample $x_i$ with the click label $y_i \in \{0,1 \}$, $\mathbf{V}_{r,i}$ and $\mathbf{V}_{t,i}$ represent the input vectors of the student and teacher model, respectively.
During training, $\mathbf{V}_{t,i}$ will be fed into the teacher model to obtain the logit $s_{t,i}$ of sample $x_i$, and then the sigmoid activation function $\sigma(\cdot)$ is adopted to compute the pCTR $\hat{p}_{t,i}$ as follows:
\begin{equation}
    \hat{p}_{t,i} = \sigma(s_{t,i})
    \label{p_ti}
\end{equation}
The standard pointwise cross-entropy (PointCE) loss is adopted to obtain a well-calibrated predicted probability of sample $x_i$:
\begin{equation}
    L_{t} = -y_i\log(\hat{p}_{t,i})-(1-y_i)\log(1-\hat{p}_{t,i})
    \label{Eq: L_t_CE}
\end{equation}

Unlike the teacher model, the student model serves online and thus only takes the non-privileged vector $\mathbf{V}_{r,i}$ as input. The output $\hat{p}_{s,i}$ of the student model is as follows:
\begin{equation}
    \hat{p}_{s,i} = \sigma(s_{s,i})
    \label{p_si}
\end{equation}
where $s_{s,i}$ is the logit of sample $x_i$, obtained from the student model before the sigmoid activation function.
We adopt two loss functions for training. The first loss is similar to the teacher model, in which the PointCE loss is adopted with the ground truth label
\begin{equation}
    L_{s,CE} = -y_i\log(\hat{p}_{s,i})-(1-y_i)\log(1-\hat{p}_{s,i})
    \label{Eq: L_s_CE}
\end{equation}
The second loss is a knowledge distillation loss $L_d(\hat{p}_{t,i},\hat{p}_{s,i})$, aiming to distill the knowledge from the teacher model. The reason is that the teacher model incorporates the privileged features and thus performs better than the student model. The final loss of the student model can be written as follows:
\begin{equation}
    L_s = \alpha L_{s,CE}(y_i,\hat{p}_{s,i}) + (1-\alpha) L_d(\hat{p}_{t,i},\hat{p}_{s,i}) 
    \label{Eq:loss-student}
\end{equation}
where $\alpha$ is the hyper-parameter that balances the importance of PointCE loss and distillation loss. The knowledge distillation loss will be discussed in detail in the following subsection.

\subsection{Listwise Privileged Features Distillation  } 
\label{3.3}
\subsubsection{Challenges}
We encounter two challenges when designing the distillation loss of the student model.
For the distillation loss, the pointwise loss is widely used~\cite{yang2022toward, xu2020privileged, liu2022position}:
\begin{equation}
    L_d = -\sigma(\frac{s_{t,i}}{\tau}) \log(\sigma(\frac{s_{s,i}}{\tau})) -(1-\sigma(\frac{s_{t,i}}{\tau})) \log(1-\sigma(\frac{s_{s,i}}{\tau}))
    \label{Eq:l_d_PointCE}
\end{equation}
where $\tau > 0$ denotes a temperature parameter. A larger $\tau$ yields a softer distribution of predictions. With $\tau \rightarrow +\infty$ and the zero-meaned assumption of logits, $L_d$ can be equivalent to MSE loss that matches the logits directly \cite{hinton2015distilling}.
\textbf{\textit{The first challenge}} is that the pointwise loss is sub-optimal when distilling the ranking ability from the teacher model.
Concretely, the pointwise loss does not consider the relative item orders for items within the same list and treats each item independently based on the i.i.d. assumption. However, data in recommendation tasks are non-i.i.d.~\cite{cao2007learning,xia2008listwise}, where the click probability of an item is affected by other items in the same list ~\cite{sheng2022joint,huang2022deep}. Therefore, it is insufficient for the pointwise loss to learn from the teacher model in CTR prediction.

Some listwise losses ~\cite{cao2007learning,qin2008query,xia2008listwise} have been proposed to optimize the model with each list as a learning instance, thus naturally accounting for the information of the relative item orders. However, although the direct utilization of typical listwise losses can improve the model's ranking ability over the pointwise one, it causes the student model's predictions to lose probabilistic meaning as pCTR, destroying the model's calibration ability. This issue is unacceptable for CTR prediction and thus poses \textbf{\textit{the second challenge}}.

\subsubsection{Calibration-compatible Listwise Distillation}
To address the above two challenges, inspired by RCR \cite{bai2023regression}, we first formally define the \textit{calibration-compatible} property of the distillation loss and analyze the calibration-compatible property of the pointwise loss and the commonly used listwise losses. We then give our solutions to design the calibration-compatible listwise distillation loss.

\textbf{Definition 1.} \textit{A distillation loss $L_d$ is calibration-compatible if it can achieve global minima when the PointCE losses of both student and teacher model achieve global minima for any candidate item $x_i$.}

We can easily prove the pointwise distillation loss is calibration-compatible. To be specific, for each list $q$, let $P_i = \mathbb{E}{[}y_i|q,x_i{]}$ be the ground truth click probability conditioned on sample $x_i$. Assume that we draw $n_i$ samples $x_i^k$ of $x_i$ from its true label distribution $Y_i$, with $y_i^k$ as the label of the $k$-th sample. We can see that the PointCE loss is minimized when $\hat{p}_{s,i} = \sum_k^{n_i} y_i^k/n_i,\hat{p}_{t,i} = \sum_k^{n_i} y_i^k/n_i$, where $\sum_k^{n_i} y_i^k/n_i = \mathbb{E}{[}y_i|q,x_i{]}$ with $n_i \rightarrow +\infty$. Therefore, the PointCE losses of the student and teacher model are calibrated and can always achieve the global minima ~\cite{yan2022scale} simultaneously when
\begin{equation}
\hat{p}_{s,i} = \hat{p}_{t,i} = P_i
    \label{Eq:PointCE_min}
\end{equation}
For the pointwise distillation loss in Eq.~\eqref{Eq:l_d_PointCE}, it is minimized when $\hat{p}_{s,i} = \hat{p}_{t,i}$ and thus obviously satisfies the calibration-compatible property. 

Take the commonly used listwise loss ListNet~\cite{cao2007learning} as an example, we prove it is not calibration-compatible when employed as distillation loss:
\begin{equation}
   L^{ListNet}_d = -\sum_{i=1}^{n} \frac{\hat{p}_{t, i}}{\sum_{j=1}^{n}\hat{p}_{t, j}} \log\frac{\exp(s_{s,i})}{\sum_{j=1}^{n} \exp(s_{s,j})}
    \label{Eq:loss-ListCE}
\end{equation}
where $n$ is the number of items for the list containing $x_i$. By the rules of differentiation, we can know that the ListNet distillation loss achieves global minima when
\begin{equation}
    \frac{\exp(s_{s,i})}{\sum_{j=1}^{n}\exp(s_{s, j})}
    =
    \frac{\hat{p}_{t,i}}{\sum_{j=1}^{n}\hat{p}_{t, j}} 
    \label{Eq:ListCE_min}
\end{equation}
From Eq.~\eqref{Eq:PointCE_min} and Eq.~\eqref{Eq:ListCE_min}, we can observe that $L^{ListNet}_d$ is not minimized (i.e., $\exp (s_{s,i}) \neq P_i$, for $i = 1\dots n$) when both the PointCE losses of the student and teacher model achieve global minima, thus not satisfying the calibration-compatible property. 

In addition, if another popular listwise loss ListMLE~\cite{xia2008listwise} is employed as distillation loss:
\begin{equation}
    L_d^{ListMLE} = -\sum_{i=1}^n\log \frac{\exp(s_{s,\pi_i})}{\sum_{j=i}^n\exp(s_{s,\pi_j})}
    \label{Eq:loss-ListMLE}
\end{equation}
where $\pi_i$ represents the sample $x_{\pi_i}$ ranked at position $i$ in the predictions generated by the teacher model, sorted in descending order.
We can observe that $L_d^{ListMLE}$ is formulated through likelihood loss that does not have a global minima during optimization, and thus it is also not calibration-compatible.
Indeed, widely used listwise losses are designed to improve the model's ranking ability without additional consideration for the calibration-compatible property when employed as the distillation loss, thus pushing the logits to fit different
objectives and corrupting the model's calibration ability.

To enable the listwise distillation loss calibration-compatible, we propose a novel privileged features distillation framework, Calibration-compatible LIstwise Distillation (CLID), which distills the ranking ability from the teacher model while preserving the student model's calibration ability.

In detail, we project $\hat{p}_{t,i}$ and $\hat{p}_{s,i}$ onto the probability simplex to form the ground-truth distribution $P_{t}(x_i)$ and the score distribution $P_{s}(x_i)$ as follows:
\begin{equation}
    P_{t}(x_i) = \frac{\hat{p}_{t, i}}{\sum_{j=1}^{n}\hat{p}_{t, j}}, P_{s}(x_i) = \frac{\hat{p}_{s, i}}{\sum_{j=1}^{n}\hat{p}_{s, j}}
\end{equation}
These probabilities encode the likelihood of sample $x_i$ appearing at the top of the ranked list. Given these two distributions, we use the cross entropy loss to penalize the difference between them and obtain the \emph{calibration-compatible listwise distillation loss}:
\begin{equation}
   L_d^{CLID} = -\sum_{i=1}^{n} P_{t}(x_i)\log P_{s}(x_i)
    \label{Eq:loss-clid}
\end{equation}

We can prove the distillation loss formulation in Eq.~\eqref{Eq:loss-clid} is calibration-compatible.
Specifically, by the rules of differentiation, the distillation loss can achieve the global minima when
\begin{equation}
    \frac{\hat{p}_{s,i}}{\sum_{j=1}^{n}\hat{p}_{s, j}} 
    =
    \frac{\hat{p}_{t,i}}{\sum_{j=1}^{n}\hat{p}_{t, j}} 
    \label{Eq:ListCE_sigma_min}
\end{equation}
From Eq. \eqref{Eq:PointCE_min} and Eq. \eqref{Eq:ListCE_sigma_min}, we can observe that when the PointCE losses of the student and teacher model achieve global minima, the distillation loss $L_d^{CLID}$ is also minimized. Therefore, this tailored distillation loss is calibration-compatible and can preserve the student model's calibration ability when distilling the teacher model's ranking ability.

\section{Experiment}
In this section, we conduct an extensive amount of experiments to understand the effectiveness of CLID over various baselines. 

\subsection{Experiment Setup}
\subsubsection{Datasets.} To demonstrate the generalization of CLID, our experiments are conducted on two widely-adopted public datasets and one production dataset. The public datasets are two popular LTR datasets, Web30K and Istella-S, with multi-graded labels ranging from 0 (irrelevant) to 4 (perfectly relevant). Following previous works ~\cite{bruch2020stochastic,yan2022scale,bai2023regression}, we binarize the labels (labels 1, 2, 3, 4 as 1, label 0 as 0) for the CTR prediction task. Since the public datasets lack the position feature, we utilize contextual features as privileged features. Specifically, in industrial recommendation systems, CTR models of the ranking and re-ranking stages are both trained with the exposed items to users. Therefore, we use the public datasets to simulate the training and serving process of the ranking stage, where the features regarding contextual documents of the candidate document in the same query are privileged features, which are inaccessible for the ranking model when testing. 
For convenience, we utilize the mean pooling operation on contextual features to obtain privileged representation vectors for all methods.  
Both public datasets contain training, validation, and test set. Final model evaluation is performed on the test set. The experiments on the public datasets are repeated for $5$ trials, and the mean metrics and $95\%$ confidence intervals are reported. For consistency, we also employ contextual features as privileged features to train the ranking model in the production dataset.

\textbf{Web30K\footnote{https://www.microsoft.com/en-us/research/project/mslr/} ~\cite{qin2013introducing}.} Web30K is a public LTR dataset, including 31,531 queries split into training, validation, and test partitions with 18,919, 6,306, and 6,306 queries, respectively.  
On average, about 119 candidate documents are associated with each query, where query-document pairs are represented by 136 numerical features and graded with a 5-level relevance label. After binarized, the percentages for labels 0 and 1 are 51.4\% and 48.6\%, respectively.

\textbf{Istella-S\footnote{http://quickrank.isti.cnr.it/istella-dataset/} ~\cite{lucchese2016post}.} 
Istella-S is also a public LTR dataset composed of 33,018 queries and 220 features representing each query-document pair. It has, on average, about 103 candidate documents associated with each query. After binarized, the percentages for labels 0 and 1 are 82.1\% and 17.9\%.

\textbf{Production.} 
The production dataset is sampled from the impression log of the Alibaba online advertising system. Specifically, the impression and click data from 2022/11/16 and 2022/11/17 are adopted for training and testing. We train the model by hours and test it in the next hour, similar to the system's online activity. The dataset consists of billions of samples with hundreds of features. 

\subsubsection{Baselines.} We utilize six baseline methods for a comprehensive comparison. We first involve recent non-PFD (PAL and PriDropOut) methods to show the generalization of CLID. We then compare the state-of-the-art PFD methods involving the mainstream pointwise-based PFD method and our extended listwise-based PFD methods with the commonly used listwise losses (ListMLE and ListNet) as distillation losses to show the superiority of CLID in both ranking and calibration performance.
The details of these baselines are listed as follows:\\
(i) \textbf{Base:} It takes the non-privileged features as input and is optimized by the PointCE loss. The Base method is the one deployed in our production system. \\
(ii) \textbf{PriDropOut ~\cite{zhang2023towards}:} It constructs a shallow tower using privileged features as input. After applying a dropout layer, its logits are added to the main tower's logits to compute pCTR during training.
During testing, the shallow tower is discarded, and pCTR is computed by the logits of the main tower. \\
(iii) \textbf{PAL ~\cite{guo2019pal}:} It calculates the pCTR by multiplying the probability score of the shallow tower with privileged features and the main tower with non-privileged features when training. During testing, only the probability score of the main tower serves as pCTR. \\
(iv) \textbf{Base+Pointwise ~\cite{xu2020privileged,liu2022position,yang2022toward}:} It employs the PFD framework to improve model performance, where the teacher model takes the non-privileged and privileged features as input. The distillation loss is the pointwise loss in Eq.~\eqref{Eq:l_d_PointCE}.\\
(v) \textbf{Base+ListMLE:} It employs the PFD framework to distill the privileged features from the teacher model, with Eq.~\eqref{Eq:loss-ListMLE} as the distillation loss.\\
(vi) \textbf{Base+ListNet:} It employs the PFD framework to distill the privileged features from the teacher model with Eq.~\eqref{Eq:loss-ListCE} as the distillation loss.\\
Our proposed \textbf{CLID} adopts Eq.~\eqref{Eq:loss-clid} as the distillation loss. 
\begin{table*}[!tb]

\renewcommand{\arraystretch}{1.1}
 
	\caption{A comparison of model performance (95\% confidence intervals) for different methods on two widely-used public datasets. The mean results are reported on the test set. The best results are highlighted in boldface. $\circ/\bullet$  indicates that CLID is statistically worse/better than the compared method by student pairwise $t$-test at 95\% confidence level.  }
 \vspace{-1em}
	\centering
 \resizebox{\textwidth}{!}{	
    \begin{tabular}{c|lll|lll}
    \hline
    	\multicolumn{1}{l|}{Datasets}  & \multicolumn{3}{c|}{Web30K} & \multicolumn{3}{c}{Istella-S} \\
		\hline
		\multicolumn{1}{l|}{Metrics}  & \multicolumn{1}{c}{NDCG@10$\uparrow$} & \multicolumn{1}{c}{LogLoss$\downarrow$} & 
		\multicolumn{1}{c|}{ECE$\downarrow$} &  \multicolumn{1}{c}{NDCG@10$\uparrow$} & \multicolumn{1}{c}{LogLoss$\downarrow$} & 
		\multicolumn{1}{c}{ECE$\downarrow$}\\
		 \hline
   \multicolumn{1}{l|}{Base} &\multicolumn{1}{l}{0.4478 ($\pm$0.0004) $\bullet$} &\multicolumn{1}{l}{0.6101 ($\pm$0.0003) $\bullet$} &\multicolumn{1}{l|}{0.1629 ($\pm$0.0003)}  & \multicolumn{1}{l}{0.6862 ($\pm$0.0015) $\bullet$} &\multicolumn{1}{l}{0.1174 ($\pm$0.0010)} &\multicolumn{1}{l}{0.0481 ($\pm$0.0010)}   \\
   \multicolumn{1}{l|}{PriDropOut} &\multicolumn{1}{l}{0.4472 ($\pm$0.0001) $\bullet$} &\multicolumn{1}{l}{0.6204 ($\pm$0.0016) $\bullet$} &\multicolumn{1}{l|}{0.1741 ($\pm$0.0009) $\bullet$}  & \multicolumn{1}{l}{0.6922 ($\pm$0.0025) $\bullet$} &\multicolumn{1}{l}{0.1314 ($\pm$0.0028) $\bullet$} &\multicolumn{1}{l}{0.0534 ($\pm$0.0027) $\bullet$}    \\
   \multicolumn{1}{l|}{PAL} &\multicolumn{1}{l}{0.4491 ($\pm$0.0003)} &\multicolumn{1}{l}{0.6631 ($\pm$0.0028) $\bullet$} &\multicolumn{1}{l|}{0.1840 ($\pm$0.0011) $\bullet$}  & \multicolumn{1}{l}{0.7070 ($\pm$0.0025) } &\multicolumn{1}{l}{0.1212 ($\pm$0.0027) $\bullet$} &\multicolumn{1}{l}{0.0465 ($\pm$0.0013) $\circ$}    \\
   \hdashline
   \multicolumn{1}{l|}{Base+Pointwise} &\multicolumn{1}{l}{0.4483 ($\pm$0.0006) $\bullet$} &\multicolumn{1}{l}{0.6095 ($\pm$0.0003) $\bullet$} &\multicolumn{1}{l|}{0.1627 ($\pm$0.0003)}  & \multicolumn{1}{l}{0.6911  ($\pm$0.0022) $\bullet$} &\multicolumn{1}{l}{\textbf{0.1129 ($\pm$0.0009) $\circ$}} &\multicolumn{1}{l}{\textbf{0.0454 ($\pm$0.0006) $\circ$}}   \\
   \multicolumn{1}{l|}{Base+ListMLE} &\multicolumn{1}{l}{0.4484 ($\pm$0.0010)} &\multicolumn{1}{l}{0.6134 ($\pm$0.0003) $\bullet$} &\multicolumn{1}{l|}{0.1627 ($\pm$0.0003)}  & \multicolumn{1}{l}{0.7021  ($\pm$0.0006) $\bullet$} &\multicolumn{1}{l}{0.1343 ($\pm$0.0003) $\bullet$} &\multicolumn{1}{l}{0.0588 ($\pm$0.0002) $\bullet$}   \\

   \multicolumn{1}{l|}{Base+ListNet} &\multicolumn{1}{l}{0.4491 ($\pm$0.0004)} &\multicolumn{1}{l}{0.6095 ($\pm$0.0001) $\bullet$} &\multicolumn{1}{l|}{0.1674 ($\pm$0.0002) $\bullet$}  & \multicolumn{1}{l}{0.6989  ($\pm$0.0006) $\bullet$} &\multicolumn{1}{l}{0.1178 ($\pm$0.0003)} &\multicolumn{1}{l}{0.0489 ($\pm$0.0012)}   \\
   \multicolumn{1}{l|}{CLID} &\multicolumn{1}{l}{\textbf{0.4495 ($\pm$0.0007)}} &\multicolumn{1}{l}{\textbf{0.6090 ($\pm$0.0002)}} &\multicolumn{1}{l|}{\textbf{0.1626 ($\pm$0.0003)}}  & \multicolumn{1}{l}{\textbf{0.7084 ($\pm$0.0019)}} &\multicolumn{1}{l}{0.1175 ($\pm$0.0008)} &\multicolumn{1}{l}{0.0489 ($\pm$0.0005)}    \\
	\hline
 
\end{tabular}
}
\label{per_pub}

\end{table*}

\subsubsection{Evaluation Metrics.} Following previous works ~\cite{bruch2020stochastic,yan2022scale,bai2023regression}, for ranking performance on the public datasets, we adopt the widely used \textbf{NDCG@10} as the evaluation metric. The higher the NDCG@10 is, the better the ranking performance is. 
As to the production dataset, we compute the \textbf{GAUC} (Group Area Under Receiver Operating Characteristic Curve) metric to evaluate the ranking performance. GAUC has shown to be more consistent with the online performance ~\cite{sheng2021one,zhou2018deep,zhu2017optimized,sheng2022joint,bian2022can} and is also the top-line metric in our production system. It can be calculated with Eq.~\eqref{GAUC}, where $U$ is the number of users, \#impression($u$) and $\text{AUC}_{u}$ are the number of impressions and AUC of the $u$-th user, respectively.
\begin{equation}
    \text{GAUC} = \frac{\sum_{u=1}^U \#impressions(u) \times \text{AUC}_u}{\sum_{u=1}^U \#impressions(u)}
    \label{GAUC}
\end{equation}

To measure the calibration performance, we employ the widely used averaged \textbf{LogLoss} for public and production datasets, following~\cite{yan2022scale,bai2023regression,chan2023capturing,gu2021real}. The LogLoss measures the sample-level calibration error, defined in Eq.~\eqref{LogLoss}, where $N$ is the number of samples.  
The expected calibration error (\textbf{ECE}) is additionally compared on public datasets. 
Following ~\cite{yan2022scale}, for the computation of ECE, we sort the documents by model predictions and divide them into $K$ bins, each containing approximately the same number of documents. It is defined in Eq.~\eqref{ECE_pub}, where $Q$ is the number of lists, $n_q$ and $n_{q,k}$ are the number of samples in list $q$ and bin $k$ of list $q$, $y_{q,k,i}$ and $\hat{p}_{q,k,i}$ represent the true label and pCTR of the $i$-th sample in the $k$-th bin of the list $q$, respectively. Here we set $K=10$.
The lower the LogLoss and ECE are, the better the performance is.

 
\begin{equation}
    \text{LogLoss} = -\frac{1}{N}\sum_{i=1}^{N}(y_i\log\hat{p}_i+(1-y_i)\log(1-\hat{p}_i))
    \label{LogLoss}
\end{equation}

\begin{equation}
    \text{ECE} = \frac{1}{Q}\sum_{q=1}^Q \frac{1}{n_q}\sum_{k=1}^K\Big|\sum_{i=1}^{n_{q,k}}(y_{q,k,i}-\hat{p}_{q,k,i})\Big|
    \label{ECE_pub}
\end{equation}


\subsubsection{Implementation Details.} We conduct experiments on public datasets using the PT-Ranking library ~\cite{yu2020ptranking}. The codes of public datasets can be found in this link.\footnote{https://www.sdu-idea.cn/codes.php?name=CLID}
 Specifically, we fixed the neural ranking model for all compared methods to be a three-layer Dense Neural Network (DNN) whose hidden layer dimensions are $1024$, $512$, and $256$. For PriDropOut and PAL, the shallow tower contains one layer with $256$ hidden units. 
We apply the input transformation $\log1p(x)=sign(x)\log(1+|x|)$, batch normalization, weight decay and dropout. 
The dropout rate in training is set to be $0.5$ and the weight decay is set to be $0.001$.
The learning rate is tuned for each method at the validation set. For distillation baselines, the hyper-parameters of the student model keep the same as the teacher model to ensure that the change in performance comes from the distillation. For the coefficient of distillation loss $\alpha$, we experiment with the weight ratio $(1-\alpha)/\alpha$ rather than a direct tuning of $\alpha$. The weight ratio is searched in the range of $\{0.001,0.01,0.1,1,10,100, 1000, 10000 \}$. 
The impact of the weight ratio is analyzed in later experiments. 
For the Base+Pointwise method, we also search $\tau$ in the range of $\{0.001,0.01,0.1,1,10,100, 1000, 10000 \}$.
To balance ranking and calibration, the weight ratio of Base+ListMLE is set to be $0.01$ on the Web30K dataset and $10$ on the Istella-S dataset. 
The weight ratio of Base+ListNet is set to be $100$ on the Web30K dataset and $10$ on the Istella-S dataset.
The weight ratio of CLID is set to be $1$ on the Web30K dataset and $100$ on the Istella-S dataset. 
The Base+Pointwise method's weight ratio and $\tau$ are $10$ and $1$ on the Web30K dataset, and $1$ and $1$ on the Istella-S dataset.
On the production dataset, we utilize \textbf{DIN} ~\cite{zhou2018deep} as the neural network structure for all methods. 

For the distillation methods on the public datasets, we first train the teacher model before distillation and then fix the converged teacher model to instruct the student model. For production scenarios, due to the huge amount of data, the teacher and student models are updated simultaneously with distillation loss.

\begin{figure}[!tbp]
\centering
\includegraphics[width=5cm]{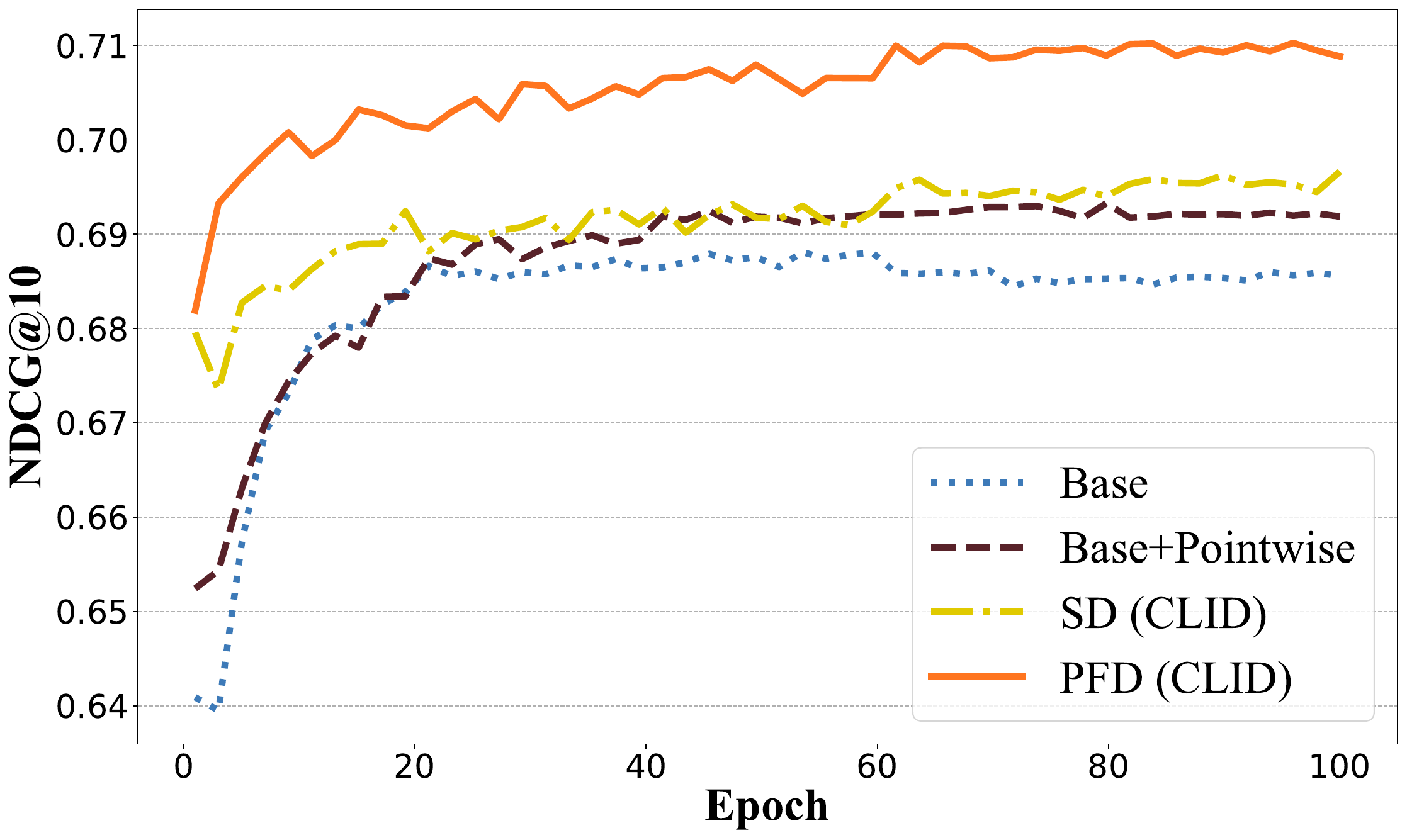}
\vspace{-1em}
\caption{A comparison of the Base, Base+Pointwise, self-distillation (SD), and privileged features distillation (PFD) methods on the Istella-S dataset. }
\label{sd_pfd}
\end{figure}
\subsection{Performance on Public Datasets} 
\subsubsection{Overall Performance.} We report results on the test set to demonstrate the superiority of CLID in both ranking and calibration performance. All methods are trained with $100$ epochs following ~\cite{yan2022scale,yang2022toward,bai2023regression}. Table \ref{per_pub} provides the model performance for all the compared methods, and the following observations can be made:\\
(i) Listwise-based PFD methods consistently outperform the pointwise-based PFD method in ranking performance. The fact demonstrates that the listwise distillation loss can account for the non-i.i.d. characteristic of data distribution, learning the teacher model's relative item order. Therefore they can distill more ranking ability from the teacher model than the pointwise one. However, the Base+ListMLE and Base+ListNet methods have compromised calibration performance. This is because the listwise distillation loss in them does not satisfy the calibration-compatible property, thus destroying the probabilistic meaning of the model's predictions as pCTR. \\
(ii) CLID achieves the best ranking performance among all compared methods while ensuring that the model's calibration performance remains equal to or surpasses that of the Base method. The observation confirms that the carefully-designed listwise distillation loss in CLID behaves well on distilling the listwise information to improve the model's ranking ability while being calibration-compatible, thereby preserving the meaning of the model's scores as pCTR. 
We further observe that the calibration of CLID is better than the Base in the Web30K dataset. We hypothesize that the improved ranking ability may further boost the calibration ability due to the calibration-compatible property. \\
(iii) The PFD methods effectively maintain the model's generalization by preserving the offline-online consistency. This is supported by the unstable ranking performance and destroyed calibration performance of the non-PFD (PAL and PriDropOut) methods. These two non-PFD methods incorporate the shallow tower at training and discard it to compute pCTR at testing, resulting in the inconsistency of pCTR between training and serving. On the one hand, the inconsistency comprises the probabilistic meaning of the model's predictions, often causing huge calibration degradation; on the other hand, it improves the ranking performance without a theoretical guarantee, such that the methods are not always effective, e.g., the degradation of ranking performance for the PriDropOut method in the Web30K dataset.

The above analyses demonstrate CLID can significantly improve the model's ranking ability while preserving its calibration ability.

\begin{figure}[!tbp]
\centering
\includegraphics[width=7.5cm]{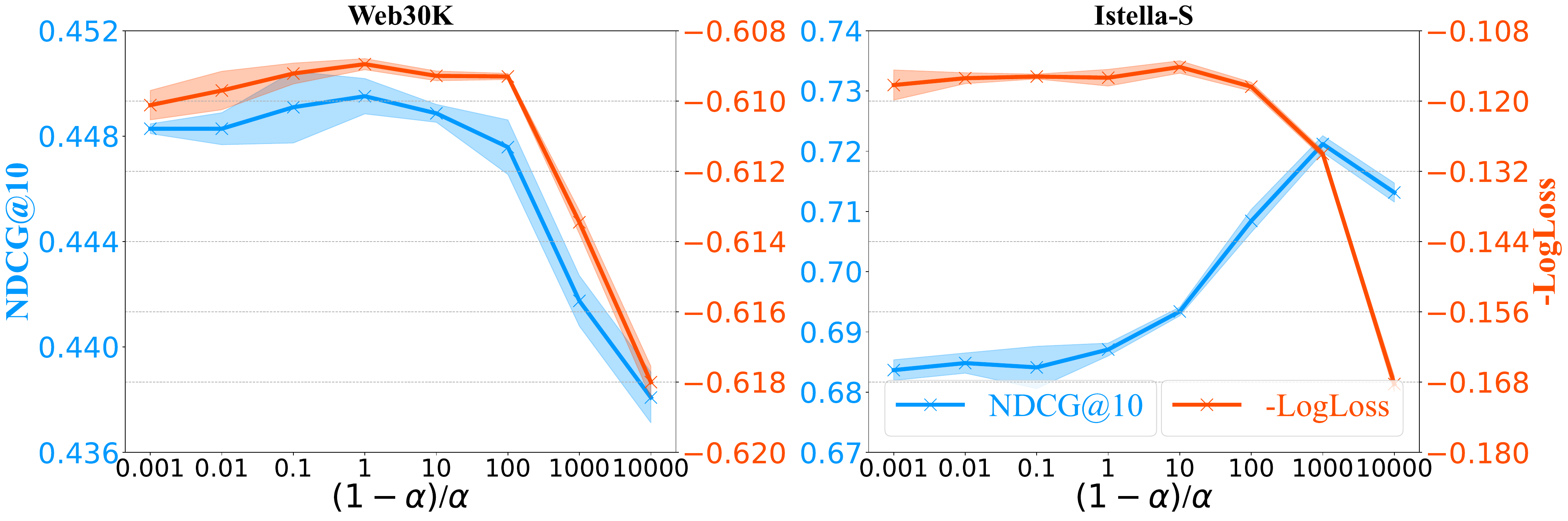}
\vspace{-1em}
\caption{The impact of weight ratio in CLID about the ranking and calibration ability on the Web30K and Istella-S datasets.}
\label{alpha_analyse}
\vspace{-2em}
\end{figure}

\subsubsection{Ablation Study} 
In this section, we aim to explore the source of CLID's effectiveness in improving ranking performance.
We hypothesize that listwise distillation loss, privileged features distillation, and knowledge distillation are essential in CLID. To verify the effect of listwise distillation loss, we utilize the Base+Pointwise method as a comparison. We adopt self-distillation (SD) as a comparison to show the effect of privileged features distillation. The teacher model of SD only takes the non-privileged features as input and uses the same listwise distillation loss as CLID. We call this method SD (CLID). Finally, Base is employed to show the effect of knowledge distillation.
We plot the NDCG@10 metric calculated on validation data for 100 epochs, and the results on the Istella-S dataset are shown in Figure. \ref{sd_pfd}.
From the figure, we can observe that all distillation methods perform better than Base, demonstrating the effectiveness of knowledge distillation.
In addition, as the training epochs increase, Base has an over-fitting problem (the performance begins to decline after $60$ epochs of training), which can be solved by knowledge distillation, as shown in the figure.
Further, we observe that PFD (CLID) achieves better performance than the Base+Pointwise method and SD (CLID) method. These two facts demonstrate the effect of listwise distillation loss and privileged features distillation, respectively.
Therefore, we can conclude that CLID's ranking improvements come from listwise distillation loss, privileged features distillation, and knowledge distillation.

\subsubsection{Impact of Weight Ratio} 
In this section, we study the impact of the weight ratio $(1-\alpha)/\alpha$ in CLID on the ranking and calibration ability. To this end, we examine the weight ratio in $\{0.001,0.01,0.1,1,10,100, 1000, 10000 \}$. The higher the weight ratio is, the more dominative the listwise distillation loss is. We plot the NDCG@10 and -LogLoss metrics evaluated on test data for each weight ratio on the Web30K and Istella-S datasets, as shown in Figure. \ref{alpha_analyse}. The higher the metric value is, the better the performance is. We observe the ranking and calibration performance gradually rise and then decrease on both datasets. We hypothesize that a large weight ratio, on the one hand, weakens the effect of PointCE loss, which uses the click label as ground truth and thus contributes significantly to the calibration; on the other hand, it makes the student model learn some noise from the teacher model and therefore cannot achieve the optimal ranking performance.
Generally, CLID can achieve a proper trade-off between the ranking and calibration performance with the weight ratio ranging from $1$ to $100$.


\begin{table}[!tb]
\renewcommand{\arraystretch}{1.1}
	\caption{A comparison of model relative improvements to the Base method for different methods in the production. The best results are highlighted in bold.}
 \vspace{-1em}
	\centering
	
    \begin{tabular}{c|lll}
         \hline
		\multicolumn{1}{l|}{Metrics}  & \multicolumn{1}{c}{GAUC$\uparrow$} & \multicolumn{1}{c}{LogLoss$\downarrow$} \\
		 \hline
   \multicolumn{1}{l|}{PriDropOut} &\multicolumn{1}{c}{+0.32\%}   &\multicolumn{1}{c}{+3.15\%}  \\
    \multicolumn{1}{l|}{PAL} &\multicolumn{1}{c}{+0.00\%}   &\multicolumn{1}{c}{+2.05\%}  \\
    \hdashline
   \multicolumn{1}{l|}{Base+Pointwise} &\multicolumn{1}{c}{+0.04\% }   &\multicolumn{1}{c}{\textbf{+0.00\%}}  \\
   \multicolumn{1}{l|}{Base+ListMLE} &\multicolumn{1}{c}{+0.37\%}   &\multicolumn{1}{c}{+0.78\%} \\
   \multicolumn{1}{l|}{Base+ListNet} &\multicolumn{1}{c}{\textbf{+0.38\%}}  &\multicolumn{1}{c}{+0.17\%} \\

   \multicolumn{1}{l|}{CLID} &\multicolumn{1}{c}{\textbf{+0.38\%}}  &\multicolumn{1}{c}{+0.02\%}  \\
	\hline
\end{tabular}
\vspace{-2em}

\label{pro_exp}
\end{table} 

\subsection{Performance on Production Dataset}
\subsubsection{Overall Performance.}
In the production dataset, we employ the methods mentioned above as baselines.
All methods are trained with one epoch following~\cite{ZhangSZJHDZ2022OneEpoch}. Note that for proprietary reasons, we only report relative numbers to our baseline (the Base method) with respect to the GAUC for ranking and LogLoss for calibration.
The results are reported in Table \ref{pro_exp}, and we can observe that the conclusions on the production dataset are aligned well with those on the public datasets:\\
(i) Listwise-based PFD methods consistently obtain significant ranking improvements over the pointwise-based PFD method, again validating the power of listwise distillation loss on learning from the teacher model. (ii) The listwise-based PFD methods cause significant calibration degradation except for CLID. This is because only the listwise distillation loss of CLID is calibration-compatible among them and thus preserving the calibration ability.  The CLID's slight increase in LogLoss might be that the added listwise loss causes the model to require more training steps for convergence than Base. (iii) PriDropOut and PAL hurt the model's generalization. To be specific, PriDropOut and PAL have huge calibration degradation due to inconsistent data distribution during training and serving. 
PriDropOut performs well, but PAL has no effect since the ranking improvements lack a theoretical guarantee.

These results again demonstrate CLID can significantly improve the model's ranking ability while preserving its calibration ability.

\section{Conclusion and Discussion}

This research studies the ways of incorporating privileged features for CTR prediction. Previous works mainly include the non-PFD and PFD methods. The former can cause training-serving inconsistency and hurt the model's generalization, while the latter adopts the pointwise distillation loss that is unsuitable for CTR prediction since it fails to consider the non-i.i.d. characteristic of the data distribution. To address these issues, we first extend the pointwise-based PFD to the listwise-based PFD and show the commonly used listwise losses can compromise the model's calibration ability. We then propose a Calibration-compatible LIstwise Distillation (CLID) method that employs a calibration-compatible listwise distillation loss to distill the teacher model's ranking ability without destroying the model's calibration ability. Both theory and experiment demonstrate the effectiveness of CLID.
\begin{acks}
This work is supported by the National Key Research and Development Program of China under Grant 2022YFC3502101, National Natural Science Foundation of China (No. 62072380 and 62272276),  Taishan Scholar Program of Shandong Province in China, and  Alibaba Group through Alibaba Innovation Research Program.
\end{acks}

\bibliographystyle{ACM-Reference-Format}

\begin{thebibliography}{61}


\ifx \showCODEN    \undefined \def \showCODEN     #1{\unskip}     \fi
\ifx \showDOI      \undefined \def \showDOI       #1{#1}\fi
\ifx \showISBNx    \undefined \def \showISBNx     #1{\unskip}     \fi
\ifx \showISBNxiii \undefined \def \showISBNxiii  #1{\unskip}     \fi
\ifx \showISSN     \undefined \def \showISSN      #1{\unskip}     \fi
\ifx \showLCCN     \undefined \def \showLCCN      #1{\unskip}     \fi
\ifx \shownote     \undefined \def \shownote      #1{#1}          \fi
\ifx \showarticletitle \undefined \def \showarticletitle #1{#1}   \fi
\ifx \showURL      \undefined \def \showURL       {\relax}        \fi
\providecommand\bibfield[2]{#2}
\providecommand\bibinfo[2]{#2}
\providecommand\natexlab[1]{#1}
\providecommand\showeprint[2][]{arXiv:#2}

\bibitem[Ai et~al\mbox{.}(2018a)]%
        {ai2018learning}
\bibfield{author}{\bibinfo{person}{Qingyao Ai}, \bibinfo{person}{Keping Bi}, \bibinfo{person}{Jiafeng Guo}, {and} \bibinfo{person}{W~Bruce Croft}.} \bibinfo{year}{2018}\natexlab{a}.
\newblock \showarticletitle{Learning a deep listwise context model for ranking refinement}. In \bibinfo{booktitle}{\emph{SIGIR}}. \bibinfo{pages}{135--144}.
\newblock


\bibitem[Ai et~al\mbox{.}(2018b)]%
        {ai2018unbiased}
\bibfield{author}{\bibinfo{person}{Qingyao Ai}, \bibinfo{person}{Keping Bi}, \bibinfo{person}{Cheng Luo}, \bibinfo{person}{Jiafeng Guo}, {and} \bibinfo{person}{W~Bruce Croft}.} \bibinfo{year}{2018}\natexlab{b}.
\newblock \showarticletitle{Unbiased learning to rank with unbiased propensity estimation}. In \bibinfo{booktitle}{\emph{SIGIR}}. \bibinfo{pages}{385--394}.
\newblock


\bibitem[Ai et~al\mbox{.}(2019)]%
        {ai2019learning}
\bibfield{author}{\bibinfo{person}{Qingyao Ai}, \bibinfo{person}{Xuanhui Wang}, \bibinfo{person}{Sebastian Bruch}, \bibinfo{person}{Nadav Golbandi}, \bibinfo{person}{Michael Bendersky}, {and} \bibinfo{person}{Marc Najork}.} \bibinfo{year}{2019}\natexlab{}.
\newblock \showarticletitle{Learning groupwise multivariate scoring functions using deep neural networks}. In \bibinfo{booktitle}{\emph{SIGIR}}. \bibinfo{pages}{85--92}.
\newblock


\bibitem[Ai et~al\mbox{.}({[n.\,d.]})]%
        {ai2021unbiased}
\bibfield{author}{\bibinfo{person}{Qingyao Ai}, \bibinfo{person}{Tao Yang}, \bibinfo{person}{Huazheng Wang}, {and} \bibinfo{person}{Jiaxin Mao}.} \bibinfo{year}{[n.\,d.]}\natexlab{}.
\newblock \showarticletitle{Unbiased learning to rank: online or offline?}
\newblock \bibinfo{journal}{\emph{TOIS}} (\bibinfo{year}{[n.\,d.]}).
\newblock


\bibitem[Bai et~al\mbox{.}(2023)]%
        {bai2023regression}
\bibfield{author}{\bibinfo{person}{Aijun Bai}, \bibinfo{person}{Rolf Jagerman}, \bibinfo{person}{Zhen Qin}, \bibinfo{person}{Le Yan}, \bibinfo{person}{Pratyush Kar}, \bibinfo{person}{Bing-Rong Lin}, \bibinfo{person}{Xuanhui Wang}, \bibinfo{person}{Michael Bendersky}, {and} \bibinfo{person}{Marc Najork}.} \bibinfo{year}{2023}\natexlab{}.
\newblock \showarticletitle{Regression Compatible Listwise Objectives for Calibrated Ranking with Binary Relevance}. In \bibinfo{booktitle}{\emph{CIKM}}. \bibinfo{pages}{4502--4508}.
\newblock


\bibitem[Bian et~al\mbox{.}(2022)]%
        {bian2022can}
\bibfield{author}{\bibinfo{person}{Weijie Bian}, \bibinfo{person}{Kailun Wu}, \bibinfo{person}{Lejian Ren}, \bibinfo{person}{Qi Pi}, \bibinfo{person}{Yujing Zhang}, \bibinfo{person}{Can Xiao}, \bibinfo{person}{Xiang-Rong Sheng}, \bibinfo{person}{Yong-Nan Zhu}, \bibinfo{person}{Zhangming Chan}, \bibinfo{person}{Na Mou}, {et~al\mbox{.}}} \bibinfo{year}{2022}\natexlab{}.
\newblock \showarticletitle{CAN: feature co-action network for click-through rate prediction}. In \bibinfo{booktitle}{\emph{WSDM}}. \bibinfo{pages}{57--65}.
\newblock


\bibitem[Bruch et~al\mbox{.}(2020)]%
        {bruch2020stochastic}
\bibfield{author}{\bibinfo{person}{Sebastian Bruch}, \bibinfo{person}{Shuguang Han}, \bibinfo{person}{Michael Bendersky}, {and} \bibinfo{person}{Marc Najork}.} \bibinfo{year}{2020}\natexlab{}.
\newblock \showarticletitle{A stochastic treatment of learning to rank scoring functions}. In \bibinfo{booktitle}{\emph{WSDM}}. \bibinfo{pages}{61--69}.
\newblock


\bibitem[Burges et~al\mbox{.}(2005)]%
        {burges2005learning}
\bibfield{author}{\bibinfo{person}{Chris Burges}, \bibinfo{person}{Tal Shaked}, \bibinfo{person}{Erin Renshaw}, \bibinfo{person}{Ari Lazier}, \bibinfo{person}{Matt Deeds}, \bibinfo{person}{Nicole Hamilton}, {and} \bibinfo{person}{Greg Hullender}.} \bibinfo{year}{2005}\natexlab{}.
\newblock \showarticletitle{Learning to rank using gradient descent}. In \bibinfo{booktitle}{\emph{ICML}}. \bibinfo{pages}{89--96}.
\newblock


\bibitem[Cao et~al\mbox{.}(2007)]%
        {cao2007learning}
\bibfield{author}{\bibinfo{person}{Zhe Cao}, \bibinfo{person}{Tao Qin}, \bibinfo{person}{Tie-Yan Liu}, \bibinfo{person}{Ming-Feng Tsai}, {and} \bibinfo{person}{Hang Li}.} \bibinfo{year}{2007}\natexlab{}.
\newblock \showarticletitle{Learning to rank: from pairwise approach to listwise approach}. In \bibinfo{booktitle}{\emph{ICML}}. \bibinfo{pages}{129--136}.
\newblock


\bibitem[Chan et~al\mbox{.}(2023)]%
        {chan2023capturing}
\bibfield{author}{\bibinfo{person}{Zhangming Chan}, \bibinfo{person}{Yu Zhang}, \bibinfo{person}{Shuguang Han}, \bibinfo{person}{Yong Bai}, \bibinfo{person}{Xiang-Rong Sheng}, \bibinfo{person}{Siyuan Lou}, \bibinfo{person}{Jiacen Hu}, \bibinfo{person}{Baolin Liu}, \bibinfo{person}{Yuning Jiang}, \bibinfo{person}{Jian Xu}, {et~al\mbox{.}}} \bibinfo{year}{2023}\natexlab{}.
\newblock \showarticletitle{Capturing Conversion Rate Fluctuation during Sales Promotions: A Novel Historical Data Reuse Approach}. In \bibinfo{booktitle}{\emph{KDD}}. \bibinfo{pages}{1--11}.
\newblock


\bibitem[Chaudhuri et~al\mbox{.}(2017)]%
        {chaudhuri2017ranking}
\bibfield{author}{\bibinfo{person}{Sougata Chaudhuri}, \bibinfo{person}{Abraham Bagherjeiran}, {and} \bibinfo{person}{James Liu}.} \bibinfo{year}{2017}\natexlab{}.
\newblock \showarticletitle{Ranking and calibrating click-attributed purchases in performance display advertising}.
\newblock In \bibinfo{booktitle}{\emph{KDD}}. \bibinfo{pages}{1--6}.
\newblock


\bibitem[Chen et~al\mbox{.}(2021)]%
        {chen2021adapting}
\bibfield{author}{\bibinfo{person}{Mouxiang Chen}, \bibinfo{person}{Chenghao Liu}, \bibinfo{person}{Jianling Sun}, {and} \bibinfo{person}{Steven~CH Hoi}.} \bibinfo{year}{2021}\natexlab{}.
\newblock \showarticletitle{Adapting interactional observation embedding for counterfactual learning to rank}. In \bibinfo{booktitle}{\emph{SIGIR}}. \bibinfo{pages}{285--294}.
\newblock


\bibitem[Chen et~al\mbox{.}(2009)]%
        {Wei2009Ranking}
\bibfield{author}{\bibinfo{person}{Wei Chen}, \bibinfo{person}{Tie{-}Yan Liu}, \bibinfo{person}{Yanyan Lan}, \bibinfo{person}{Zhiming Ma}, {and} \bibinfo{person}{Hang Li}.} \bibinfo{year}{2009}\natexlab{}.
\newblock \showarticletitle{Ranking measures and loss functions in learning to rank}. In \bibinfo{booktitle}{\emph{NeurIPS}}. \bibinfo{pages}{315--323}.
\newblock


\bibitem[Cossock and Zhang(2008)]%
        {cossock2008statistical}
\bibfield{author}{\bibinfo{person}{David Cossock} {and} \bibinfo{person}{Tong Zhang}.} \bibinfo{year}{2008}\natexlab{}.
\newblock \showarticletitle{Statistical analysis of Bayes optimal subset ranking}.
\newblock \bibinfo{journal}{\emph{IEEE Transactions on Information Theory}} \bibinfo{volume}{54}, \bibinfo{number}{11} (\bibinfo{year}{2008}), \bibinfo{pages}{5140--5154}.
\newblock


\bibitem[Fan et~al\mbox{.}(2022)]%
        {fan2022modeling}
\bibfield{author}{\bibinfo{person}{Zhifang Fan}, \bibinfo{person}{Dan Ou}, \bibinfo{person}{Yulong Gu}, \bibinfo{person}{Bairan Fu}, \bibinfo{person}{Xiang Li}, \bibinfo{person}{Wentian Bao}, \bibinfo{person}{Xin-Yu Dai}, \bibinfo{person}{Xiaoyi Zeng}, \bibinfo{person}{Tao Zhuang}, {and} \bibinfo{person}{Qingwen Liu}.} \bibinfo{year}{2022}\natexlab{}.
\newblock \showarticletitle{Modeling users' contextualized page-wise feedback for click-through rate prediction in e-commerce search}. In \bibinfo{booktitle}{\emph{WSDM}}. \bibinfo{pages}{262--270}.
\newblock


\bibitem[Gao et~al\mbox{.}(2023)]%
        {gao2023rec4ad}
\bibfield{author}{\bibinfo{person}{Jingyue Gao}, \bibinfo{person}{Shuguang Han}, \bibinfo{person}{Han Zhu}, \bibinfo{person}{Siran Yang}, \bibinfo{person}{Yuning Jiang}, \bibinfo{person}{Jian Xu}, {and} \bibinfo{person}{Bo Zheng}.} \bibinfo{year}{2023}\natexlab{}.
\newblock \showarticletitle{Rec4Ad: A Free Lunch to Mitigate Sample Selection Bias for Ads CTR Prediction in Taobao}.
\newblock \bibinfo{journal}{\emph{CIKM}} (\bibinfo{year}{2023}).
\newblock


\bibitem[Graepel et~al\mbox{.}(2000)]%
        {graepel2000large}
\bibfield{author}{\bibinfo{person}{Thore Graepel}, \bibinfo{person}{Klaus Obermayer}, {et~al\mbox{.}}} \bibinfo{year}{2000}\natexlab{}.
\newblock \showarticletitle{Large margin rank boundaries for ordinal regression}.
\newblock In \bibinfo{booktitle}{\emph{Advances in large margin classifiers}}. \bibinfo{pages}{115--132}.
\newblock


\bibitem[Gu et~al\mbox{.}(2021)]%
        {gu2021real}
\bibfield{author}{\bibinfo{person}{Siyu Gu}, \bibinfo{person}{Xiang-Rong Sheng}, \bibinfo{person}{Ying Fan}, \bibinfo{person}{Guorui Zhou}, {and} \bibinfo{person}{Xiaoqiang Zhu}.} \bibinfo{year}{2021}\natexlab{}.
\newblock \showarticletitle{Real negatives matter: continuous training with real negatives for delayed feedback modeling}. In \bibinfo{booktitle}{\emph{KDD}}. \bibinfo{pages}{2890--2898}.
\newblock


\bibitem[Guo et~al\mbox{.}(2019)]%
        {guo2019pal}
\bibfield{author}{\bibinfo{person}{Huifeng Guo}, \bibinfo{person}{Jinkai Yu}, \bibinfo{person}{Qing Liu}, \bibinfo{person}{Ruiming Tang}, {and} \bibinfo{person}{Yuzhou Zhang}.} \bibinfo{year}{2019}\natexlab{}.
\newblock \showarticletitle{PAL: a position-bias aware learning framework for CTR prediction in live recommender systems}. In \bibinfo{booktitle}{\emph{RecSys}}. \bibinfo{pages}{452--456}.
\newblock


\bibitem[Haldar et~al\mbox{.}(2020)]%
        {haldar2020improving}
\bibfield{author}{\bibinfo{person}{Malay Haldar}, \bibinfo{person}{Prashant Ramanathan}, \bibinfo{person}{Tyler Sax}, \bibinfo{person}{Mustafa Abdool}, \bibinfo{person}{Lanbo Zhang}, \bibinfo{person}{Aamir Mansawala}, \bibinfo{person}{Shulin Yang}, \bibinfo{person}{Bradley Turnbull}, {and} \bibinfo{person}{Junshuo Liao}.} \bibinfo{year}{2020}\natexlab{}.
\newblock \showarticletitle{Improving deep learning for airbnb search}. In \bibinfo{booktitle}{\emph{KDD}}. \bibinfo{pages}{2822--2830}.
\newblock


\bibitem[Han et~al\mbox{.}(2020)]%
        {han2020learning}
\bibfield{author}{\bibinfo{person}{Shuguang Han}, \bibinfo{person}{Xuanhui Wang}, \bibinfo{person}{Mike Bendersky}, {and} \bibinfo{person}{Marc Najork}.} \bibinfo{year}{2020}\natexlab{}.
\newblock \showarticletitle{Learning-to-Rank with BERT in TF-Ranking}.
\newblock \bibinfo{journal}{\emph{arXiv preprint arXiv:2004.08476}} (\bibinfo{year}{2020}).
\newblock


\bibitem[He et~al\mbox{.}(2015)]%
        {he2015delving}
\bibfield{author}{\bibinfo{person}{Kaiming He}, \bibinfo{person}{Xiangyu Zhang}, \bibinfo{person}{Shaoqing Ren}, {and} \bibinfo{person}{Jian Sun}.} \bibinfo{year}{2015}\natexlab{}.
\newblock \showarticletitle{Delving deep into rectifiers: surpassing human-level performance on imagenet classification}. In \bibinfo{booktitle}{\emph{ICCV}}. \bibinfo{pages}{1026--1034}.
\newblock


\bibitem[Hinton et~al\mbox{.}(2015)]%
        {hinton2015distilling}
\bibfield{author}{\bibinfo{person}{Geoffrey Hinton}, \bibinfo{person}{Oriol Vinyals}, \bibinfo{person}{Jeff Dean}, {et~al\mbox{.}}} \bibinfo{year}{2015}\natexlab{}.
\newblock \showarticletitle{Distilling the knowledge in a neural network}.
\newblock \bibinfo{journal}{\emph{arXiv preprint arXiv:1503.02531}} (\bibinfo{year}{2015}).
\newblock


\bibitem[Hofmann et~al\mbox{.}(2013)]%
        {hofmann2013reusing}
\bibfield{author}{\bibinfo{person}{Katja Hofmann}, \bibinfo{person}{Anne Schuth}, \bibinfo{person}{Shimon Whiteson}, {and} \bibinfo{person}{Maarten De~Rijke}.} \bibinfo{year}{2013}\natexlab{}.
\newblock \showarticletitle{Reusing historical interaction data for faster online learning to rank for IR}. In \bibinfo{booktitle}{\emph{WSDM}}. \bibinfo{pages}{183--192}.
\newblock


\bibitem[Huang et~al\mbox{.}(2022)]%
        {huang2022deep}
\bibfield{author}{\bibinfo{person}{Jianqiang Huang}, \bibinfo{person}{Xingyuan Tang}, \bibinfo{person}{Zhe Wang}, \bibinfo{person}{Shaolin Jia}, \bibinfo{person}{Yin Bai}, \bibinfo{person}{Zhiwei Liu}, \bibinfo{person}{Jia Cheng}, \bibinfo{person}{Jun Lei}, {and} \bibinfo{person}{Yan Zhang}.} \bibinfo{year}{2022}\natexlab{}.
\newblock \showarticletitle{Deep presentation bias integrated framework for CTR prediction}. In \bibinfo{booktitle}{\emph{CIKM}}. \bibinfo{pages}{4049--4053}.
\newblock


\bibitem[Lee et~al\mbox{.}(2020)]%
        {lee2020learning}
\bibfield{author}{\bibinfo{person}{Wonkyung Lee}, \bibinfo{person}{Junghyup Lee}, \bibinfo{person}{Dohyung Kim}, {and} \bibinfo{person}{Bumsub Ham}.} \bibinfo{year}{2020}\natexlab{}.
\newblock \showarticletitle{Learning with privileged information for efficient image super-resolution}. In \bibinfo{booktitle}{\emph{ECCV}}. \bibinfo{pages}{465--482}.
\newblock


\bibitem[Li et~al\mbox{.}(2007)]%
        {Li2007McRank}
\bibfield{author}{\bibinfo{person}{Ping Li}, \bibinfo{person}{Christopher J.~C. Burges}, {and} \bibinfo{person}{Qiang Wu}.} \bibinfo{year}{2007}\natexlab{}.
\newblock \showarticletitle{McRank: learning to rank using multiple classification and gradient boosting}. In \bibinfo{booktitle}{\emph{NeurIPS}}. \bibinfo{pages}{897--904}.
\newblock


\bibitem[Li et~al\mbox{.}(2023)]%
        {li2023decision}
\bibfield{author}{\bibinfo{person}{Xiang Li}, \bibinfo{person}{Shuwei Chen}, \bibinfo{person}{Jian Dong}, \bibinfo{person}{Jin Zhang}, \bibinfo{person}{Yongkang Wang}, \bibinfo{person}{Xingxing Wang}, {and} \bibinfo{person}{Dong Wang}.} \bibinfo{year}{2023}\natexlab{}.
\newblock \showarticletitle{Decision-making context interaction network for click-through rate prediction}.
\newblock \bibinfo{journal}{\emph{AAAI}} (\bibinfo{year}{2023}).
\newblock


\bibitem[Ling et~al\mbox{.}(2017)]%
        {ling2017model}
\bibfield{author}{\bibinfo{person}{Xiaoliang Ling}, \bibinfo{person}{Weiwei Deng}, \bibinfo{person}{Chen Gu}, \bibinfo{person}{Hucheng Zhou}, \bibinfo{person}{Cui Li}, {and} \bibinfo{person}{Feng Sun}.} \bibinfo{year}{2017}\natexlab{}.
\newblock \showarticletitle{Model ensemble for click prediction in bing search ads}. In \bibinfo{booktitle}{\emph{Web Conf}}. \bibinfo{pages}{689--698}.
\newblock


\bibitem[Liu et~al\mbox{.}(2022)]%
        {liu2022position}
\bibfield{author}{\bibinfo{person}{Congcong Liu}, \bibinfo{person}{Yuejiang Li}, \bibinfo{person}{Jian Zhu}, \bibinfo{person}{Fei Teng}, \bibinfo{person}{Xiwei Zhao}, \bibinfo{person}{Changping Peng}, \bibinfo{person}{Zhangang Lin}, {and} \bibinfo{person}{Jingping Shao}.} \bibinfo{year}{2022}\natexlab{}.
\newblock \showarticletitle{Position awareness modeling with knowledge distillation for CTR prediction}. In \bibinfo{booktitle}{\emph{RecSys}}. \bibinfo{pages}{562--566}.
\newblock


\bibitem[Lopez-Paz et~al\mbox{.}(2016)]%
        {lopez2016unifying}
\bibfield{author}{\bibinfo{person}{David Lopez-Paz}, \bibinfo{person}{L{\'e}on Bottou}, \bibinfo{person}{Bernhard Sch{\"o}lkopf}, {and} \bibinfo{person}{Vladimir Vapnik}.} \bibinfo{year}{2016}\natexlab{}.
\newblock \showarticletitle{Unifying distillation and privileged information}. In \bibinfo{booktitle}{\emph{ICLR}}. \bibinfo{pages}{1--12}.
\newblock


\bibitem[Lucchese et~al\mbox{.}(2016)]%
        {lucchese2016post}
\bibfield{author}{\bibinfo{person}{Claudio Lucchese}, \bibinfo{person}{Franco~Maria Nardini}, \bibinfo{person}{Salvatore Orlando}, \bibinfo{person}{Raffaele Perego}, \bibinfo{person}{Fabrizio Silvestri}, {and} \bibinfo{person}{Salvatore Trani}.} \bibinfo{year}{2016}\natexlab{}.
\newblock \showarticletitle{Post-learning optimization of tree ensembles for efficient ranking}. In \bibinfo{booktitle}{\emph{SIGIR}}. \bibinfo{pages}{949--952}.
\newblock


\bibitem[Luo et~al\mbox{.}(2018)]%
        {luo2018graph}
\bibfield{author}{\bibinfo{person}{Zelun Luo}, \bibinfo{person}{Jun-Ting Hsieh}, \bibinfo{person}{Lu Jiang}, \bibinfo{person}{Juan~Carlos Niebles}, {and} \bibinfo{person}{Li Fei-Fei}.} \bibinfo{year}{2018}\natexlab{}.
\newblock \showarticletitle{Graph distillation for action detection with privileged modalities}. In \bibinfo{booktitle}{\emph{ECCV}}. \bibinfo{pages}{166--183}.
\newblock


\bibitem[Pasumarthi et~al\mbox{.}(2020)]%
        {pasumarthi2020permutation}
\bibfield{author}{\bibinfo{person}{Rama~Kumar Pasumarthi}, \bibinfo{person}{Honglei Zhuang}, \bibinfo{person}{Xuanhui Wang}, \bibinfo{person}{Michael Bendersky}, {and} \bibinfo{person}{Marc Najork}.} \bibinfo{year}{2020}\natexlab{}.
\newblock \showarticletitle{Permutation equivariant document interaction network for neural learning to rank}. In \bibinfo{booktitle}{\emph{ICTIR}}. \bibinfo{pages}{145--148}.
\newblock


\bibitem[Pei et~al\mbox{.}(2019)]%
        {pei2019personalized}
\bibfield{author}{\bibinfo{person}{Changhua Pei}, \bibinfo{person}{Yi Zhang}, \bibinfo{person}{Yongfeng Zhang}, \bibinfo{person}{Fei Sun}, \bibinfo{person}{Xiao Lin}, \bibinfo{person}{Hanxiao Sun}, \bibinfo{person}{Jian Wu}, \bibinfo{person}{Peng Jiang}, \bibinfo{person}{Junfeng Ge}, \bibinfo{person}{Wenwu Ou}, {et~al\mbox{.}}} \bibinfo{year}{2019}\natexlab{}.
\newblock \showarticletitle{Personalized re-ranking for recommendation}. In \bibinfo{booktitle}{\emph{RecSys}}. \bibinfo{pages}{3--11}.
\newblock


\bibitem[Qin and Liu(2013)]%
        {qin2013introducing}
\bibfield{author}{\bibinfo{person}{Tao Qin} {and} \bibinfo{person}{Tie-Yan Liu}.} \bibinfo{year}{2013}\natexlab{}.
\newblock \showarticletitle{Introducing LETOR 4.0 datasets}.
\newblock \bibinfo{journal}{\emph{arXiv preprint arXiv:1306.2597}} (\bibinfo{year}{2013}).
\newblock


\bibitem[Qin et~al\mbox{.}(2010)]%
        {qin2010general}
\bibfield{author}{\bibinfo{person}{Tao Qin}, \bibinfo{person}{Tie-Yan Liu}, {and} \bibinfo{person}{Hang Li}.} \bibinfo{year}{2010}\natexlab{}.
\newblock \showarticletitle{A general approximation framework for direct optimization of information retrieval measures}.
\newblock \bibinfo{journal}{\emph{Information Retrieval}}  \bibinfo{volume}{13} (\bibinfo{year}{2010}), \bibinfo{pages}{375--397}.
\newblock


\bibitem[Qin et~al\mbox{.}(2008)]%
        {qin2008query}
\bibfield{author}{\bibinfo{person}{Tao Qin}, \bibinfo{person}{Xu-Dong Zhang}, \bibinfo{person}{Ming-Feng Tsai}, \bibinfo{person}{De-Sheng Wang}, \bibinfo{person}{Tie-Yan Liu}, {and} \bibinfo{person}{Hang Li}.} \bibinfo{year}{2008}\natexlab{}.
\newblock \showarticletitle{Query-level loss functions for information retrieval}.
\newblock \bibinfo{journal}{\emph{Information Processing \& Management}} \bibinfo{volume}{44}, \bibinfo{number}{2} (\bibinfo{year}{2008}), \bibinfo{pages}{838--855}.
\newblock


\bibitem[Sheng et~al\mbox{.}(2023)]%
        {sheng2022joint}
\bibfield{author}{\bibinfo{person}{Xiang-Rong Sheng}, \bibinfo{person}{Jingyue Gao}, \bibinfo{person}{Yueyao Cheng}, \bibinfo{person}{Siran Yang}, \bibinfo{person}{Shuguang Han}, \bibinfo{person}{Hongbo Deng}, \bibinfo{person}{Yuning Jiang}, \bibinfo{person}{Jian Xu}, {and} \bibinfo{person}{Bo Zheng}.} \bibinfo{year}{2023}\natexlab{}.
\newblock \showarticletitle{Joint optimization of ranking and calibration with contextualized hybrid model}. In \bibinfo{booktitle}{\emph{KDD}}.
\newblock


\bibitem[Sheng et~al\mbox{.}(2021)]%
        {sheng2021one}
\bibfield{author}{\bibinfo{person}{Xiang-Rong Sheng}, \bibinfo{person}{Liqin Zhao}, \bibinfo{person}{Guorui Zhou}, \bibinfo{person}{Xinyao Ding}, \bibinfo{person}{Binding Dai}, \bibinfo{person}{Qiang Luo}, \bibinfo{person}{Siran Yang}, \bibinfo{person}{Jingshan Lv}, \bibinfo{person}{Chi Zhang}, \bibinfo{person}{Hongbo Deng}, {et~al\mbox{.}}} \bibinfo{year}{2021}\natexlab{}.
\newblock \showarticletitle{One model to serve all: Star topology adaptive recommender for multi-domain ctr prediction}. In \bibinfo{booktitle}{\emph{CIKM}}. \bibinfo{pages}{4104--4113}.
\newblock


\bibitem[Sugiyama et~al\mbox{.}(2007)]%
        {SugiyamaKM2007CovariateShiftAdapt}
\bibfield{author}{\bibinfo{person}{Masashi Sugiyama}, \bibinfo{person}{Matthias Krauledat}, {and} \bibinfo{person}{Klaus{-}Robert M{\"{u}}ller}.} \bibinfo{year}{2007}\natexlab{}.
\newblock \showarticletitle{Covariate shift adaptation by importance weighted cross validation}.
\newblock \bibinfo{journal}{\emph{JMLR}}  \bibinfo{volume}{8} (\bibinfo{year}{2007}), \bibinfo{pages}{985--1005}.
\newblock


\bibitem[Swaminathan and Joachims(2015)]%
        {swaminathan2015batch}
\bibfield{author}{\bibinfo{person}{Adith Swaminathan} {and} \bibinfo{person}{Thorsten Joachims}.} \bibinfo{year}{2015}\natexlab{}.
\newblock \showarticletitle{Batch learning from logged bandit feedback through counterfactual risk minimization}.
\newblock \bibinfo{journal}{\emph{JMLR}} \bibinfo{volume}{16}, \bibinfo{number}{1} (\bibinfo{year}{2015}), \bibinfo{pages}{1731--1755}.
\newblock


\bibitem[Vapnik et~al\mbox{.}(2015)]%
        {vapnik2015learning}
\bibfield{author}{\bibinfo{person}{Vladimir Vapnik}, \bibinfo{person}{Rauf Izmailov}, {et~al\mbox{.}}} \bibinfo{year}{2015}\natexlab{}.
\newblock \showarticletitle{Learning using privileged information: similarity control and knowledge transfer.}
\newblock \bibinfo{journal}{\emph{JMLR}} \bibinfo{volume}{16}, \bibinfo{number}{1} (\bibinfo{year}{2015}), \bibinfo{pages}{2023--2049}.
\newblock


\bibitem[Wang et~al\mbox{.}(2015)]%
        {wang2015emotion}
\bibfield{author}{\bibinfo{person}{Shangfei Wang}, \bibinfo{person}{Yachen Zhu}, \bibinfo{person}{Lihua Yue}, {and} \bibinfo{person}{Qiang Ji}.} \bibinfo{year}{2015}\natexlab{}.
\newblock \showarticletitle{Emotion recognition with the help of privileged information}.
\newblock \bibinfo{journal}{\emph{IEEE Transactions on Autonomous Mental Development}} \bibinfo{volume}{7}, \bibinfo{number}{3} (\bibinfo{year}{2015}), \bibinfo{pages}{189--200}.
\newblock


\bibitem[Wang et~al\mbox{.}(2016)]%
        {wang2016learning}
\bibfield{author}{\bibinfo{person}{Xuanhui Wang}, \bibinfo{person}{Michael Bendersky}, \bibinfo{person}{Donald Metzler}, {and} \bibinfo{person}{Marc Najork}.} \bibinfo{year}{2016}\natexlab{}.
\newblock \showarticletitle{Learning to rank with selection bias in personal search}. In \bibinfo{booktitle}{\emph{SIGIR}}. \bibinfo{pages}{115--124}.
\newblock


\bibitem[Wang et~al\mbox{.}(2018)]%
        {wang2018position}
\bibfield{author}{\bibinfo{person}{Xuanhui Wang}, \bibinfo{person}{Nadav Golbandi}, \bibinfo{person}{Michael Bendersky}, \bibinfo{person}{Donald Metzler}, {and} \bibinfo{person}{Marc Najork}.} \bibinfo{year}{2018}\natexlab{}.
\newblock \showarticletitle{Position bias estimation for unbiased learning to rank in personal search}. In \bibinfo{booktitle}{\emph{WSDM}}. \bibinfo{pages}{610--618}.
\newblock


\bibitem[Wu et~al\mbox{.}(2022)]%
        {wu2022adversarial}
\bibfield{author}{\bibinfo{person}{Kailun Wu}, \bibinfo{person}{Weijie Bian}, \bibinfo{person}{Zhangming Chan}, \bibinfo{person}{Lejian Ren}, \bibinfo{person}{Shiming Xiang}, \bibinfo{person}{Shu-Guang Han}, \bibinfo{person}{Hongbo Deng}, {and} \bibinfo{person}{Bo Zheng}.} \bibinfo{year}{2022}\natexlab{}.
\newblock \showarticletitle{Adversarial gradient driven exploration for deep click-through rate prediction}. In \bibinfo{booktitle}{\emph{KDD}}. \bibinfo{pages}{2050--2058}.
\newblock


\bibitem[Xia et~al\mbox{.}(2008)]%
        {xia2008listwise}
\bibfield{author}{\bibinfo{person}{Fen Xia}, \bibinfo{person}{Tie-Yan Liu}, \bibinfo{person}{Jue Wang}, \bibinfo{person}{Wensheng Zhang}, {and} \bibinfo{person}{Hang Li}.} \bibinfo{year}{2008}\natexlab{}.
\newblock \showarticletitle{Listwise approach to learning to rank: theory and algorithm}. In \bibinfo{booktitle}{\emph{ICML}}. \bibinfo{pages}{1192--1199}.
\newblock


\bibitem[Xu et~al\mbox{.}(2020)]%
        {xu2020privileged}
\bibfield{author}{\bibinfo{person}{Chen Xu}, \bibinfo{person}{Quan Li}, \bibinfo{person}{Junfeng Ge}, \bibinfo{person}{Jinyang Gao}, \bibinfo{person}{Xiaoyong Yang}, \bibinfo{person}{Changhua Pei}, \bibinfo{person}{Fei Sun}, \bibinfo{person}{Jian Wu}, \bibinfo{person}{Hanxiao Sun}, {and} \bibinfo{person}{Wenwu Ou}.} \bibinfo{year}{2020}\natexlab{}.
\newblock \showarticletitle{Privileged features distillation at Taobao recommendations}. In \bibinfo{booktitle}{\emph{KDD}}. \bibinfo{pages}{2590--2598}.
\newblock


\bibitem[Yan et~al\mbox{.}(2022)]%
        {yan2022scale}
\bibfield{author}{\bibinfo{person}{Le Yan}, \bibinfo{person}{Zhen Qin}, \bibinfo{person}{Xuanhui Wang}, \bibinfo{person}{Michael Bendersky}, {and} \bibinfo{person}{Marc Najork}.} \bibinfo{year}{2022}\natexlab{}.
\newblock \showarticletitle{Scale calibration of deep ranking models}. In \bibinfo{booktitle}{\emph{KDD}}. \bibinfo{pages}{4300--4309}.
\newblock


\bibitem[Yang et~al\mbox{.}(2022)]%
        {yang2022toward}
\bibfield{author}{\bibinfo{person}{Shuo Yang}, \bibinfo{person}{Sujay Sanghavi}, \bibinfo{person}{Holakou Rahmanian}, \bibinfo{person}{Jan Bakus}, {and} \bibinfo{person}{SVN Vishwanathan}.} \bibinfo{year}{2022}\natexlab{}.
\newblock \showarticletitle{Toward understanding privileged features distillation in learning-to-rank}. In \bibinfo{booktitle}{\emph{NeurIPS}}. \bibinfo{pages}{1--12}.
\newblock


\bibitem[Yu(2020)]%
        {yu2020ptranking}
\bibfield{author}{\bibinfo{person}{Hai-Tao Yu}.} \bibinfo{year}{2020}\natexlab{}.
\newblock \showarticletitle{PT-ranking: A benchmarking platform for neural learning-to-rank}.
\newblock \bibinfo{journal}{\emph{arXiv preprint arXiv:2008.13368}} (\bibinfo{year}{2020}).
\newblock


\bibitem[Zhang et~al\mbox{.}(2022a)]%
        {zhang2022keep}
\bibfield{author}{\bibinfo{person}{Yujing Zhang}, \bibinfo{person}{Zhangming Chan}, \bibinfo{person}{Shuhao Xu}, \bibinfo{person}{Weijie Bian}, \bibinfo{person}{Shuguang Han}, \bibinfo{person}{Hongbo Deng}, {and} \bibinfo{person}{Bo Zheng}.} \bibinfo{year}{2022}\natexlab{a}.
\newblock \showarticletitle{KEEP: An industrial pre-training framework for online recommendation via knowledge extraction and plugging}. In \bibinfo{booktitle}{\emph{CIKM}}. \bibinfo{pages}{3684--3693}.
\newblock


\bibitem[Zhang et~al\mbox{.}(2023)]%
        {zhang2023towards}
\bibfield{author}{\bibinfo{person}{Yunan Zhang}, \bibinfo{person}{Le Yan}, \bibinfo{person}{Zhen Qin}, \bibinfo{person}{Honglei Zhuang}, \bibinfo{person}{Jiaming Shen}, \bibinfo{person}{Xuanhui Wang}, \bibinfo{person}{Michael Bendersky}, {and} \bibinfo{person}{Marc Najork}.} \bibinfo{year}{2023}\natexlab{}.
\newblock \showarticletitle{Towards Disentangling Relevance and Bias in Unbiased Learning to Rank}. In \bibinfo{booktitle}{\emph{KDD}}. \bibinfo{pages}{5618--5627}.
\newblock


\bibitem[Zhang et~al\mbox{.}(2022b)]%
        {ZhangSZJHDZ2022OneEpoch}
\bibfield{author}{\bibinfo{person}{Zhao{-}Yu Zhang}, \bibinfo{person}{Xiang{-}Rong Sheng}, \bibinfo{person}{Yujing Zhang}, \bibinfo{person}{Biye Jiang}, \bibinfo{person}{Shuguang Han}, \bibinfo{person}{Hongbo Deng}, {and} \bibinfo{person}{Bo Zheng}.} \bibinfo{year}{2022}\natexlab{b}.
\newblock \showarticletitle{Towards understanding the overfitting phenomenon of deep click-through rate models}. In \bibinfo{booktitle}{\emph{CIKM}}. \bibinfo{pages}{2671--2680}.
\newblock


\bibitem[Zhao et~al\mbox{.}(2023b)]%
        {zhao2023entire}
\bibfield{author}{\bibinfo{person}{Yunfeng Zhao}, \bibinfo{person}{Xu Yan}, \bibinfo{person}{Xiaoqiang Gui}, \bibinfo{person}{Shuguang Han}, \bibinfo{person}{Xiang-Rong Sheng}, \bibinfo{person}{Guoxian Yu}, \bibinfo{person}{Jufeng Chen}, \bibinfo{person}{Zhao Xu}, {and} \bibinfo{person}{Bo Zheng}.} \bibinfo{year}{2023}\natexlab{b}.
\newblock \showarticletitle{Entire Space Cascade Delayed Feedback Modeling for Effective Conversion Rate Prediction}. In \bibinfo{booktitle}{\emph{CIKM}}. \bibinfo{pages}{4981--4987}.
\newblock


\bibitem[Zhao et~al\mbox{.}(2023a)]%
        {zhao2023copr}
\bibfield{author}{\bibinfo{person}{Zhishan Zhao}, \bibinfo{person}{Jingyue Gao}, \bibinfo{person}{Yu Zhang}, \bibinfo{person}{Shuguang Han}, \bibinfo{person}{Siyuan Lou}, \bibinfo{person}{Xiang-Rong Sheng}, \bibinfo{person}{Zhe Wang}, \bibinfo{person}{Han Zhu}, \bibinfo{person}{Yuning Jiang}, \bibinfo{person}{Jian Xu}, {et~al\mbox{.}}} \bibinfo{year}{2023}\natexlab{a}.
\newblock \showarticletitle{COPR: Consistency-Oriented Pre-Ranking for Online Advertising}.
\newblock \bibinfo{journal}{\emph{CIKM}} (\bibinfo{year}{2023}).
\newblock


\bibitem[Zhao et~al\mbox{.}(2019)]%
        {zhao2019recommending}
\bibfield{author}{\bibinfo{person}{Zhe Zhao}, \bibinfo{person}{Lichan Hong}, \bibinfo{person}{Li Wei}, \bibinfo{person}{Jilin Chen}, \bibinfo{person}{Aniruddh Nath}, \bibinfo{person}{Shawn Andrews}, \bibinfo{person}{Aditee Kumthekar}, \bibinfo{person}{Maheswaran Sathiamoorthy}, \bibinfo{person}{Xinyang Yi}, {and} \bibinfo{person}{Ed Chi}.} \bibinfo{year}{2019}\natexlab{}.
\newblock \showarticletitle{Recommending what video to watch next: a multitask ranking system}. In \bibinfo{booktitle}{\emph{RecSys}}. \bibinfo{pages}{43--51}.
\newblock


\bibitem[Zheng et~al\mbox{.}(2022)]%
        {zheng2022cbr}
\bibfield{author}{\bibinfo{person}{Zhi Zheng}, \bibinfo{person}{Zhaopeng Qiu}, \bibinfo{person}{Tong Xu}, \bibinfo{person}{Xian Wu}, \bibinfo{person}{Xiangyu Zhao}, \bibinfo{person}{Enhong Chen}, {and} \bibinfo{person}{Hui Xiong}.} \bibinfo{year}{2022}\natexlab{}.
\newblock \showarticletitle{CBR: context bias aware recommendation for debiasing user modeling and click prediction}. In \bibinfo{booktitle}{\emph{Web Conf}}. \bibinfo{pages}{2268--2276}.
\newblock


\bibitem[Zhou et~al\mbox{.}(2018)]%
        {zhou2018deep}
\bibfield{author}{\bibinfo{person}{Guorui Zhou}, \bibinfo{person}{Xiaoqiang Zhu}, \bibinfo{person}{Chenru Song}, \bibinfo{person}{Ying Fan}, \bibinfo{person}{Han Zhu}, \bibinfo{person}{Xiao Ma}, \bibinfo{person}{Yanghui Yan}, \bibinfo{person}{Junqi Jin}, \bibinfo{person}{Han Li}, {and} \bibinfo{person}{Kun Gai}.} \bibinfo{year}{2018}\natexlab{}.
\newblock \showarticletitle{Deep interest network for click-through rate prediction}. In \bibinfo{booktitle}{\emph{KDD}}. \bibinfo{pages}{1059--1068}.
\newblock


\bibitem[Zhu et~al\mbox{.}(2017)]%
        {zhu2017optimized}
\bibfield{author}{\bibinfo{person}{Han Zhu}, \bibinfo{person}{Junqi Jin}, \bibinfo{person}{Chang Tan}, \bibinfo{person}{Fei Pan}, \bibinfo{person}{Yifan Zeng}, \bibinfo{person}{Han Li}, {and} \bibinfo{person}{Kun Gai}.} \bibinfo{year}{2017}\natexlab{}.
\newblock \showarticletitle{Optimized cost per click in taobao display advertising}. In \bibinfo{booktitle}{\emph{KDD}}. \bibinfo{pages}{2191--2200}.
\newblock


\end{thebibliography}

\balance

\end{document}